# Roadmap towards Personalized Approaches and Safety Considerations in Non-Ionizing Radiation: From Dosimetry to Therapeutic and Diagnostic Applications

**Ilkka Laakso[1]\*, Margarethus Marius Paulides[2,3], Sachiko Kodera[4], Seungyoung Ahn[5], Christopher L Brace[6], Marta Cavagnaro[7], Ji Chen[8], Zhi-De Deng[9], Valerio De Santis[10], Yinliang Diao[11], Lourdes Farrugia[12], Mauro Feliziani[10], Serena Fiocchi[13], Francesco Fioranelli[14], Takashi Hikage[15], Sergey Makaroff[16,17], Maya Mizuno[18], Alexander Opitz[19], Emma Pickwell-MacPherson[20], Punit Prakash[21], Dario B. Rodrigues[22], Kensuke Sasaki[18], Takuya Sakamoto[23], Zachary. Taylor[24], Hubregt J. Visser[25], Desmond T. B. Yeo[26], and Akimasa Hirata[4]\***

[1] Department of Electrical Engineering and Automation, Aalto University, Finland
[2] Care + Cure Lab of the Electromagnetics Group (EM4Care+Cure), Department of Electrical Engineering, Eindhoven University of Technology, Eindhoven, The Netherlands
[3] Department of Radiotherapy, Erasmus University Medical Center Cancer Institute, Rotterdam, The Netherlands
[4] Department of Electrical and Mechanical Engineering, Nagoya Institute of Technology, Japan
[5] Cho Chun Shik Graduate School of Mobility, Korea Advanced Institute of Science and Technology, Daejeon, South Korea
[6] Departments of Radiology and Biomedical Engineering, University of Wisconsin-Madison, USA
[7] Department of Information Engineering, Electronics, and Telecommunications, Sapienza University of Rome, Italy
[8] Department of Electrical and Computer Engineering, University of Houston, USA
[9] Computational Neurostimulation Research Program, Noninvasive Neuromodulation Unit, Experimental Therapeutics and Pathophysiology Branch, National Institute of Mental Health, NIH 10 Center Drive, Bethesda, MD 20892, USA.
[10] Department of Industrial and Information Engineering and Economics, University of L'Aquila, 67100 L'Aquila, Italy
[11] College of Electronic Engineering, South China Agricultural University, China
[12] Department of Physics, University of Malta, Malta
[13] CNR – National Research Council, Institute of Electronics Computer and Telecommunication (IEIIT), Italy
[14] Department of Microelectronics, Delft University of Technology, the Netherlands
[15] Faculty of Information Science and Technology, Hokkaido University, Japan
[16] Electrical and Computer Engineering Department, Worcester Polytechnic Institute, Worcester, MA 01609, USA
[17] 3Athinoula A. Martinos Ctr. for Biomedical Imaging, Massachusetts General Hospital, Harvard Medical School, Charlestown, MA 02129, USA.
[18] National Institute of Information and Communications Technology, Japan
[19] Department of Biomedical Engineering, University of Minnesota, USA
[20] Department of Physics, University of Warwick, Coventry, CV4 7AL, UK.
[21] Department of Biomedical Engineering, The George Washington University, USA
[22] Department of Radiation Oncology, University of Maryland School of Medicine, Baltimore, MD, USA

[23] Department of Electrical Engineering, Kyoto University, Japan
[24] Department of Electronics and Nanoengineering, Aalto University, Finland
[25] Department of Electrical Engineering, Eindhoven University of Technology, The Netherlands
[26] GE HealthCare, Technology and Innovation Center, Niskayuna, NY, USA

E-mail: ahirata@nitech.ac.jp, ilkka.laakso@aalto.fi

## Abstract

This roadmap provides a comprehensive and forward-looking perspective on the individualized application and safety of non-ionizing radiation (NIR) dosimetry in diagnostic and therapeutic medicine. Covering a wide range of frequencies, i.e., from low-frequency to terahertz, this document provides an overview of the current state of the art and anticipates future research needs in selected key topics of NIR-based medical applications. It also emphasizes the importance of personalized dosimetry, rigorous safety evaluation, and interdisciplinary collaboration to ensure safe and effective integration of NIR technologies in modern therapy and diagnosis.

# Contents

# 1. Introduction

Akimasa Hirata[1], Margarethus Marius Paulides[2,3], Ilkka Laakso[4]

[1] Department of Electrical and Mechanical Engineering, Nagoya Institute of Technology, Japan
[2] Care + Cure Lab of the Electromagnetics Group (EM4Care+Cure), Department of Electrical Engineering, Eindhoven University of Technology, Eindhoven, The Netherlands
[3] Department of Radiotherapy, Erasmus University Medical Center Cancer Institute, Rotterdam, The Netherlands
[4] Department of Electrical Engineering and Automation, Aalto University, Finland

The rapid advancement of non-ionizing radiation (NIR) technologies has resulted in transformative innovations in medical applications, especially in diagnostics, monitoring, and therapeutic interventions. NIR covers a wide spectrum of frequencies, ranging from static and low-frequency electromagnetic fields (EMFs) to radio frequencies, millimeter waves, terahertz, and optical radiation. Although optical techniques such as optical coherence tomography and multiphoton microscopy have been extensively studied and reviewed in recent years (Hong *et al.*, 2017; Wang *et al.*, 2022a), this roadmap focuses on non-optical NIR technologies in the terahertz and lower frequency ranges. Given the distinct physical interactions and clinical applications of these modalities, we aim to address the complementary domain of non-optical NIR technologies, which are governed by electrical stimulation and electromagnetic heating, making them suitable for non-superficial tissue diagnostics, neuromodulation, thermal therapy, and advanced sensing. Progress in NIR technologies has shown great potential for improving healthcare outcomes. In addition to advances in personalized modelling, ongoing reductions in cost and improvements in device miniaturization and flexibility have facilitated the wider adoption of systems with smaller footprints and lower operational burdens. Nevertheless, individualized approaches are often required to ensure both efficacy and safety because of the complexity of biological interaction variability in tissue properties among individuals. Digitized human body phantoms (models) enable personalized modelling in the computational domain, which requires an accurate representation of tissue properties, that is, (i) human body morphology and tissue composition and (ii) tissue physical constants (dielectric properties and thermal constants).

A foundational development in this area occurred in 1996 when a comprehensive database of the dielectric properties of different tissues was introduced by Gabriel *et al.* (1996). This was later expanded to include measurements of malignant tissues (e.g., Lazebnik *et al.* (2007)), as summarized in recent reviews by Sasaki *et al.*

(2022). Significant strides have also been made in human body modelling for safety and medical applications. For instance, Dimbylow's normalized human voxel model in 1997 (Dimbylow, 1997) based on the ICRP database, and Nagaoka's first female voxel model (Nagaoka *et al.*, 2003) were key milestones, followed by the development of a complete set of voxel models, i.e., a Virtual Family, by the IT'IS Foundation (Gosselin *et al.*, 2014). These models have since then been extended, deformed, and personalized for various applications ranging from the evaluation of electromagnetic exposure to the planning of patient-specific treatment. The development of pipelines for generating personalized human models has evolved to support specific applications such as hyperthermia treatment (Paulides *et al.*, 2010) and, more recently, brain stimulation (Saturnino *et al.*, 2019b; Laakso *et al.*, 2015).

Besides personalized human head models, including data-driven models, recent advances in computational simulations have enabled the use of subject-specific and population-based human models for detailed dosimetry analysis and population-based human models for a more sophisticated analysis of NIR–tissue interactions. These models support application-specific design and safety evaluations, while accounting for anatomical variability. These developments have led to a shift from reliance on generic phantom-based evaluations to the integration of real-time physiological monitoring based on personalized modelling. Importantly, the accuracy requirements and modelling targets differ considerably across clinical applications such as neuromodulation, thermal ablation, and implant safety (Winter *et al.*, 2021), which require tailored dosimetric strategies. Nevertheless, quantifying and controlling energy deposition at relevant spatial and temporal scales, as well as capturing individual differences in anatomy, physiology, and pathology, remain challenging. With the advancement of NIR technologies towards clinical translation and widespread adoption, establishing integrated frameworks that combine personalized anatomical modelling, real-time monitoring, and adaptive control strategies, as well as maintaining compliance with safety standards and regulatory guidelines, are important.

In addition to computational modelling and simulation, experimental dosimetry plays a key role in validating and translating models into clinically viable solutions. This includes the use of phantom measurements, magnetic resonance (MR) thermometry, and probe-based validation. Moreover, based on fundamental studies on dielectric properties, millimeter-wave (MMW) and terahertz technologies have been increasingly applied for monitoring physiological parameters in the human body.

This roadmap brings together international experts on dosimetry and associated techniques to synthesize current knowledge and outline future directions across major application areas of NIR in medicine. Each section addresses the status, key challenges, and emerging scientific and technological advances aimed at realizing effective and safe individualized NIR-based medical and healthcare applications. As these technologies approach clinical translation, it is essential to address scientific and technical challenges, as well as practical issues such as regulatory approval, interoperability with existing systems, and integration into clinical workflows. The roadmap is organized as follows.

Section 2 reviews the foundational elements of dosimetry, focusing on tissue dielectric properties and computational electromagnetic methods, both of which are crucial for accurate modelling across the low-to-radiofrequency (RF) ranges. Tissue dielectric properties vary by tissue and malignancy type, and are influenced by frequency, temperature, and thermophysiological state. Although traditional measurement techniques remain important, recent advances, including artificial intelligence-assisted modelling, magnetic resonance imaging (MRI)-based electrical property tomography, and inverse estimation methods, have enabled the noninvasive, high-resolution characterization of both healthy and pathological tissues. Despite these developments, persisting computation dosimetry challenges regarding tissue properties are tissue inhomogeneity and anisotropy, their intraindividual nature and their variability, due to e.g. thermoregulation. Emerging computational techniques, such as hybrid methods, multiscale simulations, and machine learning, have improved the accuracy and computational efficiency of electromagnetic dosimetry modelling. Further advances in simulation and measurement techniques may enable dosimetry evaluations that accurately account for real-world individual anatomy and dynamic physiological responses. This in turn supports virtual design, therapy guidance and safety assessment for a more streamlined development and uptake of a wide range of NIR-based medical technologies.

In Section 3, the roadmap highlights recent advances in transcranial electrical stimulation (TES) and transcranial magnetic stimulation (TMS), which are two key noninvasive neuromodulation techniques for the brain that use NIR in the static and low-frequency ranges. TES uses weak direct or low-frequency alternating currents to modulate brain activity using electrodes attached to the scalp. Recent efforts in individualized TES have focused on optimizing electrode placement and electric field modelling using individual MRI data; however, challenges persist due to interindividual variability in TES outcomes. In contrast, TMS directly activates neurons via high/strong electric fields induced by a pulsed magnetic field generated by a magnetic coil placed on the scalp. Recent technological innovations in TMS include digitally steerable coil arrays, multiscale neuronal simulations, and real-time individualized electric field modelling.

Section 4 analyses the dosimetry needs for thermal oncological therapies, that is, RF/microwave hyperthermia (HT) and radiofrequency/microwave ablation (RFA/MWA). HT increases tumour temperature (39–45 °C) to enhance radiation, chemotherapy, and immunotherapy efficacy. Dosimetry involves parameters, such as the electrical component of the EMF, specific absorption rate (SAR), temperature (e.g., T50, T90, $T_{max}$), and thermal dose (CEM43). Although invasive temperature probes remain standard, imaging-based thermometry is starting to enable noninvasive three-dimensional (3D) monitoring with adequate accuracy. Notably, HT was the first application that relied on human body models for device optimization, treatment planning and real-time treatment guidance. Currently, hybrid electromagnetic–thermal simulations and adaptive modelling are widely used. Yet, uncertainties in dielectric, thermal, and thermoregulation tissue properties, as well as patient positioning inside the HT device, limit simulation accuracy in patient-specific settings. Phantom-based validation and imaging-integrated systems help to address these challenges, improve clinical consistency and develop improved hybrid models. Thermal ablation techniques such as RFA and MWA achieve direct tissue destruction at higher temperatures (>50 °C). RFA exploits electrical currents with the heating constrained by the dependency on thermal conduction, whereas MWA enables deeper and faster heating due to the greater contribution of electromagnetic energy absorption to growth of the ablation zone. In contrast to HT, ablation has a considerably larger therapeutic temperature range, thus the dosimetry problem is geared towards assessing whether a threshold thermal dose is attained in targeted tissue, while restricting temperatures in non-targeted regions below safety limits. In routine clinical practice, dosimetry is typically limited to localized temperature sensor data and post-treatment imaging to assess extent of the ablation zone. Medical imaging-based thermometry and computational models are under development to improve intraprocedural feedback. Variations in tissue perfusion and other biophysical properties, both across patients and dynamically during an ablation, can lead to variable outcomes. Efforts continue to enhance real-time monitoring and model-guided planning to improve precision and treatment outcomes for both HT and ablation.

In Section 5, RF-based diagnostic technologies are examined across the following three frequency domains: RF (employed in MRI), MMW sensing, and emerging THz applications. For ultra-high-field parallel transmission MRI, rapid and accurate prediction of local SAR remains a critical challenge. Recent advances in personalized human models and machine-learning-enabled local SAR estimation demonstrated promise in improving safety and performance. A remaining key challenge includes estimation of tissue dielectrics and local thermal dose in real time. In MMW sensing, radar-based techniques provide noncontact, privacy-respecting alternatives to conventional methods for monitoring respiration, cardiac activity, and body movement. Nonetheless, there are challenges such as individual variability, multidimensional signal processing, and the need to ensure robustness under real-world conditions. The combination of machine learning and MMW sensors paves the way for their clinical deployment. THz sensing, which is still in its early development, shows promise for noninvasive hydration measurement and analysis of the protein structure in the skin and eyes. Customized imaging systems and portable scanners have been developed for ophthalmological and dermatological applications. Across all modalities, personalized modelling and/or measurement innovations are essential to achieve safe, effective, and clinically viable NIR-based diagnostics.

In Section 6, the roadmap addresses the following three issues related to implantable medical devices (IMDs): wireless power transfer (WPT), safety of IMDs during MRI, and malfunction of IMDs interacting with EMF. WPT for IMDs emphasizes their potential to replace conventional batteries and to reduce the need for surgical interventions. Current WPT systems span near-, mid-, and far-field approaches, and vary in range, efficiency, and power depending on the device application. These variations lead to challenges for efficient power delivery to deep implants, EMF safety, interference management, and biocompatibility. In parallel, MRI safety and electromagnetic compatibility (EMC) evaluations remain critical, especially for active implants exposed to RF and gradient fields. Patient-specific modelling and high-fidelity electromagnetic simulations improve the precision of safety evaluations.

By this roadmap of the needs for dosimetry in seminal applications, we hope to provide a good overview of the key challenges and future directions for the individualized and safe application of NIR in medicine and healthcare. The authors hope that this will serve as a valuable reference for researchers and practitioners in this evolving field, and hence that will accelerate the development and increase the impact of tailored NIR-based medical applications.

**Acknowledgement:**

This roadmap article was organized by A.H., I.L., and M.P. The authors of each section are listed in the headings of the corresponding section. The document was edited and prepared by A.H., I.L., and S. K. After the document was prepared, the authors reviewed the corresponding sections and approved the final version. The summary of each section has been reviewed by the authors of each section. A.H. received funding from JSPS KAKENHI 21H04956, which was partly related to this article. M.P. was financially supported by Eurostars project 3868 (SENS-THERM) and project P21-34 (CARES) of the research programme Perspectief, which is financed by the Dutch Research Council (NWO).

## 2. Methods for personalized and accurate NIR dosimetry

### 2.1 Tissue dielectric properties

Kensuke Sasaki[1] and Lourdes Farrugia[2]

1 National Institute of Information and Communications Technology, Japan

2 Department of Physics, University of Malta, Malta

Status.

Tissue dielectric properties (i.e., permittivity and electrical conductivity) are fundamental physical constants that characterize the electrical response of biological materials. They are essential for electromagnetic modelling used across various research fields, such as nonionizing radiation (NIR) safety, medical/healthcare applications, and information and communication applications via the human body. The values of these tissue dielectric properties are both frequency- and temperature-dependent and vary with the tissue type and pathology. In the case of NIR dosimetry for the human body, dielectric measurements for each tissue and frequency have been performed; review studies have been reported by Foster (1989), Gabriel *et al.* (1996), Amin *et al.* (2019), and Sasaki *et al.* (2022). Because tissue dielectric properties are sensitive to tissue conditions (e.g., temperature and hydration), research in this area also explores potential applications in medical diagnostics and therapeutic modalities.

Current and Future Challenges.

NIR dosimetry covers a broad frequency range, from extremely low frequencies (approximately tens of hertz) to millimetre-wave frequencies (Hirata *et al.*, 2021). With ongoing advancements in wireless communication, particularly beyond 5G and 6G technologies, there is growing interest in submillimetre frequencies extending up to a few terahertz (THz). This introduces new research opportunities and challenges for characterizing the dielectric properties of biological tissues across a wider spectrum. Contrary to advances in high-resolution computational models, comprehensive dielectric property data below 1 MHz remain scarce and represent a key challenge for future research. Furthermore, investigating the dielectric response at extremely low frequencies (1–300 Hz) is also of interest, particularly for understanding the dielectric dispersion mechanisms in tissues.

The characteristics of tissue dielectric properties for medical and healthcare applications are topics of interest in the frequency range from tens of megahertz to a few gigahertz. Research in this field requires accurate knowledge of the dielectric properties of healthy and malignant tissues for diagnostic purposes and the mechanism by which the dielectric properties change in response to various pathological conditions. Furthermore, therapeutic technologies such as hyperthermia and ablation rely on the thermal and dielectric properties of tissues, which vary owing to tissue degeneration and temperature changes during treatment. Although the time-variant nature of tissue dielectric properties is known, an efficient dosimetry method that accounts for this nature has not yet been fully developed.

At higher frequencies, i.e., millimetre- and submillimetre-wave frequencies, the penetration depth of electromagnetic waves becomes shallow. This makes the dielectric properties of superficial tissues (tissues that constitute the skin and eye) a special topic of interest. THz time-domain spectroscopy (THz-TDS) has extended the frequency range over which the tissue dielectric properties can be measured (Pickwell *et al.*, 2004). Future research should address the dominant factors (e.g., skin wrinkles and hydration are the potential factors) that influence the tissue dielectric properties, and the modelling of the dielectric properties, which embed the dominant factors as parameters, is essential, particularly in NIR dosimetry.

Advances in science and technology to meet the challenges.

Dielectric measurement techniques have been developed over time, enabling the characterization of tissue dielectric properties over a broad frequency spectrum. At kilohertz frequencies or lower, the primary method is to use dielectric measurement cells or sensors constructed using pair(s) of electrodes (Schwan and Kay, 1957). For higher frequencies, typically beyond hundreds of megahertz, coaxial line-based sensors are a common approach, as pioneered by Stuchly and Stuchly (1980). These conventional measurement techniques remain fundamental and are widely used because they enable broadband frequency measurements, thus demonstrating their effectiveness.

Measurement guidelines have been developed to ensure proper usage and measurement reproducibility when using such techniques (Farrugia *et al.*, 2024).

Recent advancements in the determination of tissue dielectric properties apply artificial intelligence (AI) techniques to conventional measurement techniques (Bonello *et al.*, 2020) and magnetic resonance imaging technology (Mandija *et al.*, 2019). The imaging approach resolves electromagnetic inverse problems and allows noninvasive estimation of tissue dielectric properties with the resolution of imaging technology. More recently, an innovative approach has been proposed to fuse computational dosimetry techniques and conventional imaging techniques to estimate dielectric properties (Kangasmaa and Laakso, 2022). These advancements in accuracy and efficiency allow for the understanding of intratissue variability that are difficult to capture using conventional measurement techniques.

These methodologies continue to advance, paving the way for deeper insights into the variability of dielectric properties within the same tissue type. Conventional measurement techniques often struggle to quantify such variations, and in most methodologies, average dielectric properties are measured. However, modern AI-driven and computational imaging approaches have the potential to accumulate precise, high-resolution data on how dielectric constants fluctuate due to physiological and pathological changes. These advancements have enabled more detailed, personalized assessments of tissue dielectric properties, which could enhance applications in medical diagnostics, treatment planning, and NIR safety evaluations.

Concluding remarks

For the tissue dielectric properties, which are fundamental physical quantities essential for computational NIR dosimetry, the demand for data according to frequency and tissue type (including tissue state) tends to expand with the development of science and technology. Because exhaustive measurements of all tissue types and frequencies are impractical, the development of advanced estimation approaches can become a core topic in the near future. In parallel, the continuous accumulation of high-quality experimental data by scientifically well-characterized measurement technology and ensuring its traceability are essential in validating estimation methods and remain vital for ensuring traceable and reliable dosimetry.

## 2.2 Low-frequency and radiofrequency computational methods

Sachiko Kodera[1] and Yinliang Diao[2]

1 Department of Electrical and Mechanical Engineering, Nagoya Institute of Technology, Japan

2 College of Electronic Engineering, South China Agricultural University, China

Status

**Low-Frequency (LF) Dosimetry:** Accurate modelling of the interaction between LF electric and magnetic fields and biological tissues is essential for applications such as bioelectric signal analysis, neuromodulation, nerve stimulation, and electrotherapy. Many living cells and tissues generate and respond to electrical signals, and the dominant mechanism of LF EMF exposure involves the stimulation of excitable tissues, such as nerves and muscles. Consequently, dosimetry focuses on induced electric fields, whether generated by external stimuli or biological activity. LF dosimetry is based on a quasistatic approximation in which the EMF wavelengths are significantly larger than the dimensions of the system under study.

Computational models are essential for optimizing electrotherapy techniques. These models include transcranial electrical stimulation (TES), transcranial magnetic stimulation (TMS), spinal cord stimulation, and source localization for electroencephalography (EEG) and electrocardiography (ECG). Anatomical human models commonly used in dosimetric analyses are constructed *via* tissue segmentation based on medical images, mainly using magnetic resonance imaging (MRI) and computed tomography. For medical applications, such as brain stimulation, procedures have been developed to enable semiautomatic segmentation from individual MRI scans to provide personalized EMF dosimetry (Laakso *et al.*, 2015; Saturnino *et al.*, 2019b).

The boundary element method (BEM) (Sauter and Schwab, 2010) and finite element method (FEM) (Wang and Eisenberg, 1994) are widely used in electrophysiological source reconstruction techniques (e.g., EEG, MEG, and ECG) and brain stimulation techniques (e.g., TES and TMS). The BEM calculates the potential of surface elements at the interface between tissue compartments. This is preferred when the surface potential and current density are the primary quantities of interest. However, the BEM cannot handle inhomogeneous and anisotropic tissues (Dayarian and Khadem, 2024). In the FEM, the entire volume is discretized into small elements (typically tetrahedrons), and the potential at all nodes is calculated. The FEM is particularly advantageous for incorporating arbitrary geometries and heterogeneous and anisotropic tissue properties. However, the computational resources are higher, and singularities are encountered when using point dipoles as current sources in EEG forward models (Beltrachini, 2018).

Unlike mesh-based methods, the scalar-potential finite-difference (SPFD) method (Dawson and Stuchly, 1996) uses voxel-based uniform grids, which makes it suitable for anatomical models generated from medical images. The SPFD method is effective for heterogeneous and anisotropic models and has been widely used in numerical dosimetry, for evaluating LF EMF safety. Compared to the FEM, it usually requires less computational memory and time (Soldati and Laakso, 2020). However, one major problems with voxel-based modelling is the computational artifacts arising from the staircasing approximation of curved tissue interfaces (Hirata *et al.*, 2021). Other approaches are the quasistatic finite-difference time-domain (FDTD) method (Gandhi and Chen, 1992), which uses a step function to approximate the voltage source, and the impedance method (Deford and Gandhi, 1985), which represents human tissue as a network of impedances.

**Radiofrequency (RF) Dosimetry:** In RF-based medical applications, such as hyperthermia therapy, thermal ablation, and MRI-related safety assessments, dosimetry is essential to estimate power absorption in biological tissues and the corresponding temperature rise. The specific absorption rate is a key parameter for evaluating the temperature increase in tissues, and it affects treatment and safety considerations. Numerical dosimetry for RF exposure often uses the FDTD method (Taflove and Hagness, 2003) because of its ability to handle inhomogeneous tissues and geometric complexities, as well as its suitability for parallel computation using graphics processing units. Similar to LF dosimetry, the FEM and BEM are also widely used in RF applications. The FEM is beneficial for applications involving complex vascular geometries, such as tumour thermal therapy, although it typically requires higher computational resources. However, it is also used in cases in which computational efficiency is a priority; though, it suffers from scalability issues when applied to complex and inhomogeneous geometries (Hall and Hao 2012). Thermal modelling is essential for evaluating tissue heating for RF-based medical treatments. The Pennes bioheat transfer equation (BHTE) (Pennes, 1948) is widely used to estimate the temperature rise in the human body. The BHTE was originally proposed as a simplified equation for heat transfer between tissues, but it has been extended to take into account other factors such as evaporation, blood perfusion, convection, and temperature changes in arterial and venous blood (Fiala *et al.*, 2001; Tucci *et al.*, 2021; Radmilović-Radjenović *et al.*, 2022). Due to these extensions, the BHTE has become applicable to various medical applications, including RF-induced heating in patients within MRI systems (Ertürk *et al.*, 2015) and thermal therapy (Keangin *et al.*, 2011; Paulides *et al.*, 2013).

Challenges and Future Directions.

**LF Dosimetry Challenges:** Accurate computational dosimetry is essential for enabling individual dosimetry for diagnosis and treatment planning, thereby ensuring the safety of individuals exposed to EMF and establishing safe exposure limits for various scenarios (Diao *et al.*, 2023). To achieve this, continuous improvements in computational methods and the development of more accurate human models considering tissue anisotropy are required. The fast computations are also important, especially when real-time feedback is required. Multiscale modelling combining EMF computations and nerve activation models (Reilly and Diamant, 2011) has attracted attention. This approach can help us better understand the effects of LF EMF at cellular level. Furthermore, exploring the effects of non-sinusoidal waveforms on nerve interactions is critical because it provides deeper insights into safety guidelines and informs them of therapeutic strategies.

**RF Dosimetry Challenges:** In hyperthermia therapy, personalized dosimetry has attracted attention because of its potential to optimize patient-specific treatment outcomes (Verhaart *et al.*, 2015; Wells *et al.*, 2021). This trend is likely to be extended to other RF treatments. However, the accurate modelling of tissue structures, such as blood vessels and tumours, remains a challenge (Lu *et al.*, 2024). The variation in the dielectric properties further complicates the dosimetry accuracy in hyperthermia and thermal ablation treatments. In RF dosimetry for thermal ablation, the temperature dependence of the dielectric properties of biological tissues should be clarified (Rossmann and Haemmerich, 2014). Exposure assessments should be standardized to ensure consistency in medical and safety applications. A challenge in thermal modelling is the limited availability of reliable data on the thermal properties of biological tissues. Tissue thermal properties were partly estimated by extrapolation from *in vitro* or animal data or empirical equations (Duck, 1990; Mcintosh and Anderson, 2010; Hasgall *et al.*, 2023). However, differences in water content among individuals affect not only dielectric properties but also thermal properties (Michel *et al.*, 2017). To improve model accuracy, integrating subject-specific data on thermal properties, including individual variations, is essential. Another key challenge is modelling thermoregulatory responses (Kodera *et al.*, 2018). Most computational studies do not consider the dynamic blood flow and temperature feedback mechanisms, which are critical for realistic temperature increase predictions (Fiala *et al.*, 2001). Integrating more detailed data on thermoregulatory responses is essential for improving the accuracy of the temperature increase estimation and for assessing temperature increases considering individual variabilities.

## Advances in Science and Technology to Meet Challenges

Dosimetry can help us better understand the interactions between EMFs and the human body and guide the development of devices that use these interactions. Advancements in dosimetry are likely to emerge alongside new device designs and applications, combining sophisticated multiphysics simulations for design and optimization with swift prototyping and testing tools.

**Advances in LF dosimetry**: Innovations in computational methods, such as improved FEM algorithms and hybrid techniques, promise greater accuracy and efficiency in modelling the interactions between EMFs and the human body. For instance, subtraction methods have been proposed to address singularity issues in the FEM (Beltrachini, 2018). The hybrid BEM–FEM method, which combines the advantages of both approaches by ensuring continuity of the current density across the interface between the isotropic domain (using the BEM) and the anisotropic domain (using the FEM), has been proposed (Dodig *et al.*, 2021). However, this approach generally requires longer simulation times. Recently, a BEM with fast multipole acceleration was proposed to speed up computations using high-resolution bioelectromagnetic models (Makarov *et al.*, 2018b). For the real-time computation of the TMS-induced electric field in the brain, a reciprocity-based method was introduced (Stenroos and Koponen, 2019). In the finite-difference method, the geometric multigrid (Laakso and Hirata, 2012) and algebraic multigrid (Stroka *et al.*, 2024) methods are used to accelerate computation, particularly when integrated with multiple graphics processing units (Xiong *et al.*, 2015).

Regarding computational artifacts, postprocessing methods that adjust percentile values based on the statistical distribution of the top 1% of electric field strengths have been proposed (Gomez-Tames *et al.*, 2018; Arduino *et al.*, 2020). In Laakso and Hirata (2012), a preprocessing smoothing conductivity algorithm was effective in reducing staircasing artifacts. However, this approach results in the loss of anatomical model details. Furthermore, recent studies have proposed an effective tensor conductance model to reduce staircasing artifacts in the SPFD method (Diao *et al.*, 2022; Diao *et al.*, 2023). Moreover, multiscale simulation techniques that incorporate more realistic nerve models are expected to improve the accuracy of threshold estimation, which would also be valuable for refining spatial averaging methods for induced electric fields.

AI, particularly deep learning, has recently seen rapid growth due to advancements in hardware and software. Studies have reported the feasibility of using deep learning for the fast estimation of TMS-induced electric fields, achieving an estimation time of <0.1 s (Yokota *et al.*, 2019; Moser *et al.*, 2024). However, large datasets are required for training, and the neural network architecture must be tailored to specific problems. Embedding physics in deep learning networks can improve interpretability and generalizability; however, challenges remain (Guo *et al.*, 2023).

**Advances in RF dosimetry**: In the field of EMF safety, high-resolution human models has been developed to replicated the detailed morphology of the skin (e.g., sweat ducts, nerves, epidermis, dermis, and subcutaneous tissue) and eye (e.g., cornea, iris, lens, sclera, vitreous body, and anterior chamber) have been developed to enhance high-precision dosimetry (Sasaki *et al.*, 2017; Haider *et al.*, 2021; Karampatzakis and Samaras, 2010). These advances could be useful for medical applications such as thermal and retinal laser therapies. One of the key challenges in hyperthermia and ablation include improving the focus of the irradiated area and minimizing the residual heat in the area surrounding the tumour (Hassan *et al.*, 2021). Achieving this requires detailed and personalized EMF dosimetry that reproduces the complex structure of the tumour and surrounding tissue, statistical approaches that consider the variability of dielectric properties, and estimation of the time course of the temperature rise considering thermoregulation. Thermoregulation modelling could be further improved by clarifying the relationship between temperature, blood flow, and sweating using measured data obtained from the human subject experiments. A more accurate understanding of these physiological responses would improve the accuracy of estimations of temperature changes in biological tissues during treatments, especially under high-intensity exposure conditions. Furthermore, considering the interindividual variability of thermal properties such as specific heat capacity could help quantify uncertainties in estimated temperature rise. Future research should focus on individual differences in thermoregulatory responses, including those related to age, sex, environmental conditions, and pathology/diseases. The use of real-time physiological monitoring data obtained from wearable sensors may enable the validation of these models more simply and under various conditions. The advanced imaging techniques such as dynamic contrast-enhanced MRI combined with computational simulations may improve the depiction of microvascular blood flow, ultimately improving the accuracy of the evaluation of thermal therapy (Singh, 2024).

Concluding Remarks

Integrating LF and RF dosimetry into medical applications is important for optimizing the therapeutic outcomes and patient safety. Although computational modelling techniques have substantially improved the accuracy and resolution of dosimetric assessments, challenges related to physiological variations, tissue properties, and thermoregulation remain an active area of research. Accurate and individualized dosimetry is expected to improve the optimization of therapeutic techniques. Improved dosimetric models can enhance the precision of these treatments and lead to improved patient outcomes. Future advances in numerical modelling combined with machine learning and real-time physiological monitoring will further enhance personalized treatment planning. Furthermore, incorporating electromagnetic interaction modelling that accounts for tissue anisotropy and dynamic physiological responses will enable dosimetry evaluations that reflect real-world exposure scenarios and individual differences, such as age, sex, and morphology. These advancements represent safer and more effective medical therapies involving EMF exposure, thus benefiting a wide range of clinical applications.

Acknowledgements.

Sachiko Kodera was supported by JSPS KAKENHI Grant Number JP24K00868.

# 3. Brain Stimulation

## 3.1 Transcranial Electrical Stimulation

Ilkka Laakso[1], Serena Fiocchi[2], Alexander Opitz[3]

[1] Department of Electrical Engineering and Automation, Aalto University, Finland

[2] CNR – National Research Council, Institute of Electronics Computer and Telecommunication (IEIIT), Italy

[3] Department of Biomedical Engineering, University of Minnesota, USA

Status

Transcranial electrical stimulation (TES) refers to a class of noninvasive neuromodulation techniques that apply weak electrical currents to the scalp to modulate brain activity. The applied current generates an electric field within the brain, which is thought to modulate the function of the cerebral cortex, without directly triggering action potentials. These methods have gained widespread attention in both research and clinical settings due to their potential applications in the treatment of various neurological and psychiatric disorders (Kuo et al., 2014).

Various forms of TES exist, each with specific mechanisms and potential advantages. The most commonly used technique is transcranial direct current stimulation (tDCS), in which a constant, low-intensity direct current is applied to alter cortical excitability depending on the current polarity (Nitsche and Paulus, 2000). Another common technique is transcranial alternating current stimulation (tACS) (Antal et al., 2008), which applies sinusoidal currents at specific frequencies to influence neural oscillations. In temporal interference stimulation (Grossman et al., 2017; Violante et al., 2023), kilohertz tACS at different frequencies is applied through two or more pairs of electrodes to produce patterns of two interfering electric fields, which allows for modulation of deeper brain structures compared to conventional tACS.

Personalized dosimetry in TES relies on advanced computational and imaging tools to optimize stimulation parameters for individual subjects. Finite Element Method (FEM)-based models, often constructed using high-resolution structural magnetic resonance images and diffusion-weighted imaging, allow for subject-specific electric field estimations by incorporating variations in skull conductivity, cerebrospinal fluid (CSF) distribution, and brain tissue properties. Freely available software platforms like SimNIBS (Saturnino *et al.*, 2019b) and ROAST (Huang et al., 2019) provide automated pipelines for generating individualized electric field simulations, enabling clinicians and researchers to predict stimulation intensity at target brain regions.

These models have allowed new strategies for tuning the electric field dose through optimization of TES electrode montages. A multitude of possible electrode placements, using either classical bipolar montage or arrays of smaller electrodes, permit shaping current flow patterns through the head or targeted stimulation of cortical areas. In this context, electrode montage personalization has emerged as approach to enhance stimulation focality and ensure consistent electric field delivery across subjects (Evans et al., 2020). Moreover, advanced strategies, such as focal tDCS setups and multi-electrode configurations (Kuo *et al.*, 2013; Niemann *et al.*, 2024; Parazzini *et al.*, 2017), have been investigated to optimize stimulation precision. More recently, MRI-free approach using EEG coordinate-based electrode placement has demonstrated to enhance both the intensity and focality of electric fields compared to conventional configurations (Caulfield and George, 2022).

Challenges and Future Directions

Personalization of TES holds the potential to significantly improve the consistency and predictability of treatment outcomes by addressing the substantial interindividual variability observed in response to TES (e.g., (López-Alonso et al., 2014; Wiethoff et al., 2014; Therrien-Blanchet et al., 2023). While the underlying causes of the variability are not yet established, several factors are thought to contribute to it, including individual head anatomy, brain features, ongoing brain state, and the resulting electric field distribution (Van Hoornweder et al., 2023). By tailoring stimulation parameters to an individual's specific characteristics, personalized TES aims to optimize the electric field strength and focality in the targeted brain regions (Simula et al., 2022; Van Hoornweder et al., 2022; Van Hoornweder et al., 2023).

However, linking individual electric fields to neurophysiological outcomes has been challenging so far. For tDCS of the primary motor cortex, studies have examined the relationship between individual electric fields and motor evoked potential size modulations, but results have been inconsistent. They have reported negative effects (Laakso et al., 2019), positive effects (Mosayebi-Samani et al., 2021), no effects (Ahn and Fröhlich, 2021), and participant-specific, nonlinear effects (Laakso et al., 2024). Individual electric field strengths have also been associated with positive effects on GABA modulation and increases in sensorimotor network strength (Antonenko et al., 2019). While prospective personalized dosing of motor cortical tDCS and tACS has been explored for reaction time and working memory, the results showed little to no improvement over conventional dosing (Joshi et al., 2023). Studies in other brain regions also show mixed results: some found strong positive associations between electric fields and functional connectivity (Indahlastari et al., 2021), while others reported weaker or more variable effects (Müller et al., 2023). Higher electric field strength was related to faster reaction times in one of two working memory tasks but had no effect heart rate variability (Razza et al., 2024a; Razza et al., 2024b). Over the visual cortex, individual tACS electric fields were associated with stronger alpha-power of the magnetoencephalogram (Kasten et al., 2019). Overall, the dose-response relationship is still unclear due to conflicting results in various studies.

Furthermore, electric field simulations are only an approximation of physical reality and include various sources of uncertainty. Initially spherical models were used to predict TES electric fields (Miranda, 2013; Miranda et al., 2009). While being only a coarse representation of the intricate head and brain morphology, spherical models nevertheless provided important insights into TES physical mechanisms. Today, anatomically realistic head models can be derived from individual MR images allowing more precise estimation of electric fields. For TES applications, electrodes are often placed based on the EEG 10-20 system (Antal et al., 2017; Wischnewski et al., 2023) which can introduce consistent variability in placement with respect to anatomical targets. With the increasing popularity of take-home TES (Palm et al., 2018) where stimulation electrodes are placed by participants themselves, localization errors are likely larger. Imprecise electrode location can have large effects on the electric fields in the brain (Opitz et al., 2018) and lead to suboptimal outcomes.

Additionally, tissue conductivity variations can introduce further uncertainty and affect the predicted electric field strength. While the directionality of predicted electric fields is generally more robust, electric field strength is more affected by conductivity uncertainties (Saturnino et al., 2018). For electric field simulations, tissue conductivities serve as input parameters, typically derived from ex vivo animal models (S. Gabriel et al., 1996), with considerable variability reported across measurements (C. Gabriel et al., 1996). Also, tissue conductivities can change during aging (Akhtari et al., 2002). For TES, non-brain conductivities, such as of the skin, skull and CSF, play a crucial role in determining the amount of current shunting occurring before the current reaches the brain (Narayan et al., 2023; Opitz et al., 2015; Salvador et al., 2012), thus affecting the electric field strength. Related to the influence of tissue conductivity uncertainties is the effect of segmentation errors. If voxel tissues are assigned an incorrect segmentation label, the conductivity distribution in the head model is altered, leading to inaccurate local electric field estimation.

## Science and Technological Advances Addressing the Challenges

Several scientific advances are needed to address the optimal electric field dose and its relationship with the response, which currently remain unclear, hindering personalization of TES. Presently available dose-response studies in volunteer participants are highly heterogeneous, having investigated various outcomes, applied different experimental protocols, variable modelling parameters, and dosing parameters for characterizing the individual electric fields (Van Hoornweder et al., 2023). To address this heterogeneity, larger studies with standardized experimental protocols and consistent electric field dose measures are needed. Furthermore, computational and experimental strategies need to be developed to avoid errors caused by inaccurate electrode localization, and in vivo measurement of electrical conductivities in humans are needed to reduce uncertainties associated with tissue electrical properties.

The majority of TES studies have not reported individual electric field data but instead report the dose in terms of electrode size and montage and stimulation current intensity, resulting in variability in electric field distributions in the brain. This makes it challenging to compare outcomes across studies and establishing dose-response relationships. However, variations in stimulation parameters will result in different induced electric fields which can be accounted for using computational modelling, allowing for meta-analyses that compare and aggregate studies based on estimated TES electric field distributions. For instance, combining electric field modelling with a traditional meta-analysis has been shown to identify brain regions involved in performance changes in working memory following tDCS (Wischnewski et al., 2024a, 2021) as well as identify different brain regions causally altered by theta and gamma (tACS) (Wischnewski et al., 2024b). This Meta-dosimetry approach could be a powerful tool to identify effective electric field dosing parameters across an increasingly growing TES literature.

In addition to computational approaches, new experimental studies in animals can provide detailed and accurate information into the TES dose-response relationship. Animal studies leveraging invasive physiological recordings have already demonstrated that TES physiological effects are dose-dependent (Alekseichuk et al., 2022; Johnson et al., 2020; Krause et al., 2022). For example, Johnson et. al (2020) showed that increasing tACS electric field strength leads to greater neuronal spike entrainment to external oscillations. Thus, recordings in animals can help identify TES parameters to achieve a desired physiological response. However, care has to be taken to translate findings in animals to human applications due to differences in head/brain anatomy leading to differences in TES electric field strengths (Alekseichuk et al., 2019).

The difficulty in identifying the TES dose response relationship may also be related to the dependence of TES effects on the brain state when stimulation is delivered. A recently developed neural network (Zhao et al., 2024) shows that the effects of tACS depend on the ongoing network oscillation and the applied stimulation frequency. For strong ongoing brain oscillations, tACS first disrupts existing spike field coupling, and only at higher intensities does it start to entrain spiking activity to the external oscillations. Additionally, real-time EEG-informed current steering has emerged as a potential approach to enhance targeting precision by adjusting TES in response to ongoing brain activity (Wischnewski et al., 2024). These findings show that ongoing brain states will affect TES dose-response relationships, and it is beneficial to consider these effects in future studies on dose response.

Concluding Remarks

Personalized computational dosimetry is a potential approach to reduce variability and improve the therapeutic effectiveness of TES. However, the optimal dosing of the brain electric fields is still unclear and there are several sources of dosimetric uncertainty. To overcome these challenges, we propose reducing the heterogeneity in study design, using standardized dosing parameters, systematic review and meta-analysis of TES literature to identify relevant dosing parameters, animal experiments on dose response, and considering brain-state dependence of TES in experimental design.

*3.2 Transcranial Magnetic Stimulation*

Zhi-De Deng[1] and Sergey N. Makaroff[2,3]

[1]Computational Neurostimulation Research Program, Noninvasive Neuromodulation Unit, Experimental Therapeutics and Pathophysiology Branch, National Institute of Mental Health, NIH 10 Center Drive, Bethesda, MD 20892, USA.

[2]Electrical and Computer Engineering Department, Worcester Polytechnic Institute, Worcester, MA 01609, USA.

[3]Athinoula A. Martinos Ctr. for Biomedical Imaging, Massachusetts General Hospital, Harvard Medical School, Charlestown, MA 02129, USA.

Status

Transcranial magnetic stimulation (TMS) is a noninvasive brain stimulation technique that uses rapidly alternating magnetic fields to induce electric fields within targeted brain regions. These electric fields are sufficiently intense to depolarize neuronal membranes and induce action potentials, enabling TMS to directly activate neurons in a focal, spatially constrained manner (Barker *et al.*, 1985). Unlike transcranial electrical stimulation (TES), such as transcranial direct current stimulation (tDCS) or transcranial alternating current stimulation (tACS), which modulates brain activity more diffusely and at subthreshold intensities, TMS achieves direct cortical activation, making it a more potent tool for probing and modulating brain function.

Since its first demonstration in the mid-1980s, TMS has evolved from a laboratory tool into a clinically validated intervention, particularly in the treatment of neuropsychiatric disorders. Its capacity to modulate neural activity in a site-specific, frequency-dependent manner has led to regulatory approvals for a range of neuropsychiatric disorders such as major depressive disorder, obsessive-compulsive disorder, migraine, and smoking cessation (Cohen *et al.*, 2022; O'Reardon *et al.*, 2007; Carmi *et al.*, 2019; Lefaucheur *et al.*, 2020). It has also become a valuable adjunct in stroke rehabilitation and a tool for cognitive enhancement (Fisicaro *et al.*, 2019; Luber and Deng, 2017; Luber and Lisanby, 2014). The appeal of TMS lies in its versatility. As both a therapeutic modality and an investigative probe of brain function, TMS occupies a unique position within the broader field of neuromodulation, bridging basic neuroscience, clinical application, and personalized medicine.

Over the past two decades, several distinct forms of TMS have emerged. Among these modalities, repetitive TMS (rTMS) has traditionally been the most established and extensively studied, forming the basis of many clinical protocols. rTMS involves delivering a series of pulses at regular intervals and has demonstrated frequency-dependent effects on cortical excitability. Low-frequency rTMS, typically administered at 1 Hz, is generally associated with inhibitory outcomes, thought to reflect mechanisms akin to long-term depression (LTD). In contrast, high-frequency rTMS, often delivered at 5 Hz or higher, tends to facilitate cortical excitability, likely through mechanisms resembling long-term potentiation (LTP)(Fitzgerald *et al.*, 2006). A notable evolution of rTMS is theta-burst stimulation (TBS), which capitalizes on brief, high-frequency bursts delivered in theta-range temporal patterns. Intermittent TBS (iTBS) tends to enhance excitability, whereas continuous TBS (cTBS) has been shown to reduce it (Huang *et al.*, 2005). TBS protocols can induce lasting plasticity with considerably shorter administration times compared to rTMS, offering logistical advantages in both research and clinical settings.

Beyond these therapeutic paradigms, single-pulse and paired-pulse TMS are commonly employed as investigative tools in neurophysiology. Single-pulse TMS can be used to assess motor threshold and corticospinal integrity. Paired-pulse paradigms—where two stimuli are delivered in close succession—allow researchers to probe intracortical facilitation and inhibition (Kujirai *et al.*, 1993). These techniques have yielded critical insights into cortical circuitry, inhibitory/excitatory balance, and the neurophysiological basis of various brain disorders.

Additionally, more experimental paradigms such as quadri-pulse stimulation (QPS) (Hamada *et al.*, 2007) and patterned burst protocols are under investigation for their capacity to induce highly specific and sustained neuroplastic changes. As stimulation technologies advance, the repertoire of TMS protocols continues to grow, offering increasingly nuanced control over how, where, and when neuromodulation occurs.

The effects of TMS unfold across multiple spatial and temporal scales, encompassing both local neurophysiological effects and network-level effects. At the site of stimulation, TMS can modulate neuronal excitability via mechanisms that resemble classical synaptic plasticity, including LTP and LTD mechanisms. These effects are believed to arise from activity-dependent modulation of synaptic strength, as well as from alterations in the balance of excitatory and inhibitory neurotransmitter systems—most notably glutamate and GABA (Ziemann *et al.*, 2008).

Moreover, the direction and magnitude of these effects are strongly influenced by stimulation parameters, such as pulse frequency, intensity, and pattern. While the local impact of TMS is important, its true functional reach emerges through distributed network modulation. Neuroimaging studies employing functional MRI (fMRI), electroencephalography (EEG), and magnetoencephalography (MEG) consistently show that TMS induces changes in functionally connected brain regions. These network-level responses are particularly evident when stimulating cortical hubs that anchor large-scale networks, such as the default mode network, salience network, or frontoparietal control network (Fox *et al.*, 2012; Eldaief *et al.*, 2011). For example, targeting the dorsolateral prefrontal cortex can modulate activity not only in the prefrontal region itself, but also in the anterior cingulate, striatum, and thalamus—regions implicated in emotional regulation and executive function. Crucially, these network effects are not merely epiphenomenal. They appear to play a central role in the therapeutic efficacy of TMS, with mounting evidence suggesting that normalization of aberrant connectivity may underlie symptom improvement in conditions such as depression. Thus, understanding TMS as a tool that reshapes neural communication—rather than simply exciting or inhibiting individual nodes—offers a more accurate and powerful framework for both research and clinical application.

Challenges and Future Directions

A major limitation of present TMS systems lies in the rigidity of coil geometry and placement during stimulation. Most coils are designed with fixed coil windings and require physical repositioning to target different brain regions. As a result, they are poorly suited for aligning stimulation with the rapidly evolving spatial-temporal dynamics of brain activity. This mismatch becomes especially critical when attempting to engage distributed or state-dependent networks, as commonly seen in higher-order cognitive tasks or neuropsychiatric conditions involving fluctuating brain states. Even protocols that target "hubs" of large-scale networks assume static topographies and lack the flexibility to adapt stimulation trajectories within or across sessions. These constraints underscore the need for systems capable of dynamic, multifocal, and state-dependent stimulation. Compounding the hardware limitations are acoustic artifacts. A persistent challenge with traditional TMS coils is their acoustic output, which often generate sound pressure levels exceeding 120–140 dB(Z) (Koponen et al., 2020). Not only does this necessitate hearing protection, but it may also inadvertently engage auditory pathways, potentially confounding both therapeutic outcomes and neurophysiological measurements.

Subthreshold and low-field TMS presents another conceptual and mechanistic challenge. Growing evidence suggests that even weak, non-depolarizing fields can modulate neural excitability, oscillatory dynamics, and plasticity (Romei *et al.*, 2016; Zmeykina *et al.*, 2020). However, the mechanisms underpinning these effects remain poorly understood, especially given that standard modelling approaches are typically tuned to threshold-level depolarization.

Alongside interest in subthreshold stimulation, the development of portable TMS systems raises both exciting opportunities and unique technical hurdles. These systems promise to increase access to TMS by enabling home-based treatments or ambulatory neuromodulation. However, portability often entails compromises in coil design, field strength, and power requirements. Integrating lightweight and energy-efficient components without sacrificing efficacy or safety remains a formidable engineering challenge. Moreover, the clinical use of portable TMS will demand new approaches to remote supervision, personalized calibration, and long-term adherence—all of which require coordination between hardware, software, and regulatory frameworks.

Finally, across both traditional and emerging applications, computational modelling remains a pivotal but underdeveloped tool. Most current frameworks terminate at the biophysical interface—predicting electric field strength or estimating neural activation thresholds (Siebner *et al.*, 2022)—without resolving how stimulation sequences influence neural circuits over time. As a result, they offer limited insight into how specific stimulation sequences shape neuronal population activity or how repeated pulses reorganize network function over time. Yet, such mechanistic understanding is critical for advancing beyond heuristic protocol design. Equally lacking are models that incorporate plasticity mechanisms capable of capturing the longer-term, cumulative effects of stimulation. While empirical findings suggest that certain TMS protocols induce LTP- or LTD-like changes (Ziemann *et al.*, 2008), these effects are typically inferred post hoc rather than predicted a priori. Biophysically grounded models that simulate synaptic plasticity (Robinson, 2011; Fung *et al.*, 2013)—particularly in response to temporally patterned input—could help identify optimal treatment frequency, session duration, or cumulative dosing strategies for clinical applications. In this way, computational modelling could evolve from a tool for anatomical targeting into a predictive framework for both mechanistic insight and individualized protocol development.

Scientific and Technological Advances Addressing the Challenges

A significant frontier in TMS coil technology is the development of multi-locus and electronically steerable stimulation systems, which aim to overcome a long-standing limitation: the need to physically reposition the coil to stimulate different cortical targets. These systems use arrays of overlapping coils whose outputs can be digitally controlled to steer the electric field in real time across the cortical surface without mechanical movement. This innovation enables stimulation of multiple brain areas in rapid succession or along complex spatial trajectories, thereby facilitating more flexible, network-oriented neuromodulation (Koponen *et al.*, 2018; Rissanen *et al.*, 2023; Nieminen *et al.*, 2022). In research contexts, multicoil arrays can even be combined with neuroimaging to allow for causal mapping of functional connectivity (Navarro de Lara *et al.*, 2021), while in clinical settings it may enable more efficient engagement of distributed circuits implicated in neuropsychiatric disorders. As such, these systems bring TMS closer to the ideals of adaptive, precision-guided neuromodulation, aligning stimulation not just to anatomy, but to dynamically shifting brain states and network configurations. Another recent advancement in coil engineering focuses not on the shape or depth of stimulation, but on the acoustic features of TMS. A novel class of quiet TMS (qTMS) coils—such as the double-containment coil (DCC)—has been developed to substantially reduce acoustic emissions (Koponen *et al.*, 2021). These coils incorporate structural isolation between the coil windings and the outer housing, dampening mechanical vibrations that give rise to the characteristic "click" of stimulation. The result is a reduction in sound levels by 18–41 dB(Z) at matched stimulation strength. Beyond enhancing comfort and safety, qTMS technology holds promise in research settings where auditory contamination of neural signals (e.g., in EEG or fMRI studies) must be minimized to avoid confounds.

One recent development in TMS hardware is a wearable, battery-powered systems capable of delivering suprathreshold repetitive stimulation. A notable example is *rTMS-tiny*, a lightweight (3 kg), fully portable rTMS system powered by a high-efficiency magnetic stimulator and an innovative bent double-T magnetic core coil (Qi *et al.*, 2025a). This system achieves E-field intensities and repetition frequencies comparable to commercial devices, while reducing power consumption to just 10% of conventional rTMS systems. Critically, *rTMS-tiny* supports delivery of 10 Hz rTMS and intermittent theta burst stimulation (iTBS) during free behaviors, enabling real-time neuromodulation in ambulatory and naturalistic settings. The ability to stimulate during movement—such as walking—has revealed state-dependent enhancement of cortical excitability, illustrating the potential for closed-loop or context-aware neuromodulation outside laboratory constraints. These breakthroughs not only address long-standing limitations in access and ecological validity, but also open new frontiers for longitudinal intervention, home-based therapy, and real-world causal neuroscience.

Complementing efforts in miniaturization and mobility, recent progress in subthreshold magnetic stimulation has expanded the therapeutic landscape even further. Recent work by Dufor et al. (2019) demonstrated that low-intensity magnetic stimulation can induce axon outgrowth and synaptogenesis—not through direct depolarization, but via biologically relevant stimulation patterns that engage cryptochrome-dependent intracellular signaling (Dufor *et al.*, 2019). These plasticity effects were critically dependent on temporal structure rather than raw field strength, underscoring the role of stimulation pattern in governing efficacy. Crucially, these findings reveal that magnetic fields themselves—independent of the induced electric field—can initiate gene regulatory cascades essential for circuit repair. In addition to pre-clinical mechanistic studies, a real-world clinical study by Jensen et al. (2025) found that low-intensity transcranial pulsating electromagnetic fields (T-PEMF), a form of subthreshold stimulation, significantly reduced depressive symptoms in patients with treatment-resistant depression following an 8-week self-administered course (Jensen *et al.*, 2025). Notably, the therapeutic benefit was most pronounced in individuals who received the intervention within two years of episode onset, suggesting a window of heightened neuroplasticity during which subthreshold approaches may be especially effective. Together, these findings suggest that subthreshold TMS engages biologically meaningful, noncanonical mechanisms of plasticity and holds promise as a low-burden, home-deliverable therapeutic option distinct from conventional high-intensity protocols.

Personalization in TMS has become an essential component of both research and clinical applications, aiming to maximize efficacy by tailoring stimulation to individual brain anatomy. A cornerstone of this personalization effort lies in electromagnetic modelling—particularly under the quasistatic approximation—which enables prediction of the electric fields induced in brain tissue during stimulation. This approximation, rigorously justified and widely adopted (Wang *et al.*, 2024), underpins most modern modelling strategies and is foundational to determining which neuronal populations are likely to be activated by a given TMS pulse. The finite element method (FEM) remains the most prevalent approach for macroscopic modelling of TMS fields. As a powerful numerical technique, FEM excels at solving the complex geometries and heterogeneous conductivity profiles inherent to the human head. Among the open-source tools available, SimNIBS stands out as the leading FEM-based platform for individualized electric field modelling (Thielscher *et al.*, 2015; Saturnino *et al.*, 2019a; Saturnino *et al.*, 2019c). It offers a streamlined pipeline for generating subject-specific head models from MRI, running simulations, and visualizing the induced electric fields.

An alternative to FEM is the boundary element method with fast multipole acceleration (BEM-FMM), a modernized variant of a technique long used in bioelectromagnetics (Makarov *et al.*, 2018a; Makarov *et al.*, 2020). BEM-FMM accelerates calculations by leveraging surface integrals, thus avoiding the volumetric meshing required in FEM. This approach proves especially advantageous in high-resolution simulations or when computational resources are limited. Its use in TMS modelling has expanded in recent years due to improvements in numerical stability and toolkits tailored to neuroscience applications.

While FEM and BEM-FMM capture how electric fields distribute across cortical and subcortical structures, they do not directly address the neural response to stimulation. This gap has motivated the development of multiscale neuronal models that couple macroscopic field simulations with biophysically detailed neuron models (Aberra *et al.*, 2020a; Aberra *et al.*, 2018; Weise *et al.*, 2023; Qi *et al.*, 2025b). These frameworks track the effects of TMS pulses on individual neurons—including action potential initiation and subthreshold membrane dynamics—and can be extended to study network-level recruitment. By resolving how fields interact with morphologically and directionally sensitive neural elements, multiscale models help bridge the divide between electromagnetic physics and neurophysiological outcome.

Recent work highlights how individualized E-field modelling is transforming TMS coil placement from a heuristic practice into a data-driven optimization problem. As reviewed by Dannhauer *et al.* (2024), subject-specific modelling pipelines now support precise dosing by accounting for factors like scalp-cortex distance, gyral geometry, and coil orientation. These pipelines, often built on platforms such as SimNIBS and the Targeting and Analysis Pipeline (TAP) (Dannhauer *et al.*, 2022), integrate E-field simulations with structural and functional imaging to inform and verify coil placement. By leveraging electromagnetic reciprocity and auxiliary dipole methods, they enable efficient optimization without sacrificing physical realism. Empirical studies have demonstrated the clinical relevance of such approaches. In a study of depressed adolescents, Deng et al. compared four common TMS targeting strategies: the 5 cm rule, Beam F3 method, MRI-guided targeting, and a computationally optimized approach. E-field modelling revealed substantial differences in target engagement, with the computational method delivering the strongest and most focal stimulation to the left dorsolateral prefrontal cortex (L-DLPFC). Moreover, among participants who completed treatment, median E-field magnitude at the L-DLPFC was linearly correlated with symptom improvement, underscoring the potential for model-informed dosing to enhance therapeutic outcomes (Deng *et al.*, 2023). These findings support broader integration of modelling pipelines into clinical workflows for personalized TMS.

Recent advances in real-time E-field modelling are pushing personalized TMS into the realm of dynamic, intra-session optimization. Traditional modelling approaches, though accurate, are computationally intensive and not practical for real-time clinical feedback. Addressing this gap, Li et al. introduced a self-supervised deep learning framework capable of predicting high-resolution E-fields within seconds, bypassing the need for time-consuming partial differential equation solutions (Li *et al.*, 2022). Trained on MRI-derived head models, the network achieved accuracies comparable to FEM solutions across realistic anatomical conditions, dramatically accelerating E-field estimation. Complementing this, Hasan et al. developed a GPU-accelerated solver that integrates reciprocity and Huygens' principles to compute E-fields in under 4 milliseconds for full-head models (Hasan *et al.*, 2025). By projecting coil-induced fields onto a reduced basis of cortical modes, their method maintains sub-3% error rates even at high spatial resolution. Importantly, these solutions are compatible with neuronavigation systems, enabling potential real-time updates to coil placement and intensity based on the evolving brain state. Together, these innovations demonstrate how modern computational strategies—ranging from physics-informed deep learning to high-efficiency reciprocal solvers—are reshaping the possibilities for adaptive, model-informed TMS delivery.

## Concluding Remarks

TMS has matured into a versatile platform for both clinical intervention and neuroscientific discovery. Yet, its full potential is only beginning to be realized. As advances in device engineering, pulse sequencing, and computational modelling converge, TMS is moving from a one-size-fits-all tool towards a precision-guided neuromodulation system. The integration of personalized electric field modelling—particularly with real-time capabilities—ushers in the possibility of adaptive TMS, where stimulation parameters are dynamically tailored to a patient's unique brain anatomy, functional state, and clinical trajectory. Multiscale modelling continues to enhance our mechanistic understanding, connecting physics with physiology, while novel coil technologies and digitally steerable arrays promise unprecedented spatial and temporal control. In the coming years, the coupling of TMS with neuroimaging, electrophysiology, and machine learning may yield closed-loop stimulation paradigms that are both individualized and responsive, accelerating our capacity to treat complex brain disorders and to interrogate the neural substrates of cognition with causal precision.

## Acknowledgments

ZDD and SNM are inventors of patents and patent applications on electrical and magnetic brain stimulation technology. ZDD is supported by the NIMH Intramural Research Program (ZIAMH002955). SNM is supported by the NIMH grant R01MH130490 and NIBIB grant R01EB035484.

# 4. Thermal Therapeutic Techniques

## 4.1 Radiofrequency and Microwave Hyperthermia

Margarethus Marius Paulides[1,2] and Dario B. Rodrigues[3]

[1] Care & Cure Lab of the Electromagnetics Group (EM4Care+Cure), Department of Electrical Engineering, Eindhoven University of Technology, Eindhoven, The Netherlands

[2] Department of Radiotherapy, Erasmus University Medical Center Cancer Institute, Rotterdam, The Netherlands

[3] Department of Radiation Oncology, University of Maryland School of Medicine, Baltimore, MD, USA

Status

### A. Background

Hyperthermia therapy (HT) has been long recognized as an effective adjuvant approach in cancer treatment. This therapy involves elevating tumour temperature to 39-45 °C for one hour, enhancing the efficacy of radiation therapy (RT), chemotherapy (CT), immunotherapy, or a combination of these modalities (Datta *et al.*, 2015). The radio-sensitization is mostly achieved via increased blood perfusion and oxygenation as well as inhibition of radiation-induced DNA damage repair (Elming *et al.*, 2019). The benefits from HT on CT relate to increased blood perfusion in tumours accommodated with an increased permeability of the tumour blood vessels, thus improving the leakage of drugs where in the heated region (Issels, 2008). Several studies also showed that hyperthermia boosts local and systemic immune response and enhance the effects of immunotherapies (Hurwitz, 2019).

In this section, we will analyse dosimetry for HT by electromagnetic (EM) techniques involving RF and microwave (MW) applicators, but the same principles are valid for conductive heating, ultrasound-based HT applicators or HT induced by nanoparticles (Kok *et al.*, 2020). HT can be applied invasively or noninvasively using externally applied power. RF/MW approaches can be further divided into local or regional HT. Local HT can be applied to both superficial and deep-seated tumours by external, intraluminal or interstitial MW methods. Regional hyperthermia is delivered using capacitive (8-27 MHz) or RF phased-array applicators (>65 MHz) to treat deep-seated pelvic or abdominal tumours. Capacitive systems provide significant unwanted fat heating and little control due to the long RF wavelengths in tissue, while RF phased arrays enable steering of a heating region that has the shape of an ellipsoid in homogenous medium. The key difference between local and regional HT approaches is the operating frequency: localized heating is achieved with MW (> 300 MHz) and deep regional heating is achieved with RF (≤300 MHz) frequencies (Rodrigues *et al.*, 2022).

In the following sections, we will provide an overview of key HT dosimetric parameters, discuss current methodologies used for their assessment, and highlight the state-of-the-art techniques along with the main challenges faced in HT dosimetry.

### B. Hyperthermia dosimetric parameters

A clear-cut relationship between thermometric parameters with treatment outcome is highly attractive because it would improve our understanding of tumour-specific mechanisms of interaction between HT and RT, CT and immunotherapy. HT dosimetric parameters can be divided in terms of the induced EM electric field component (E, units V/m), resulting temperature (T, units °C) from the EM power absorption (SAR with units W/kg), and thermal dose, a parameter that considers both temperature and heating time. SAR-based parameters include maximum SAR ($SAR_{max}$) and $SAR_{50}$, corresponding to the 50th percentile SAR distribution in tissue or phantom. Temperature-based parameters include maximum ($T_{max}$), minimum ($T_{min}$), and average $T_{ave}$ target temperature over the treatment duration, as well as $T_{90}$, $T_{50}$, and $T_{10}$, with $T_i$ representing the ith percentile temperature distribution in the target. The most common parameter for thermal dose is the thermal iso-effect dose or cumulative equivalent minutes at 43 °C (CEM43) calculated as follows:

$$CEM43 = \sum_{i=1}^{n} t_i R^{43-T_i} \tag{1}$$

where $t_i$ is the ith time interval (min), R accommodates the different rates of cell death above and below the 43 °C breakpoint in the Arrhenius plot (R = 1/4 for T < 43 °C and R = 1/2 for T ≥ 43 °C), and T is the average tissue temperature (°C) over the treatment interval $t_i$ at that measurement point (Dewhirst *et al.*, 2003; Dewhirst *et al.*, 2005; Sherar *et al.*, 1997b; Lee *et al.*, 1998; Ohguri *et al.*, 2018).

Since thermal dose distributions within a tumour volume are heterogenous, the calculation of tumour dose involves the thermal dose accumulated at each point with $T_{90}$, resulting in CEM43$T_{90}$ (Dewey, 2009; Dewhirst *et* *al.*, 2003). CEM43 parameters were derived from cell death studies as a function of temperature, but they do not include the radio- or chemo-sensitization effect. Nonetheless, there is a strong correlation between thermal dose and clinical outcomes as has been shown by several clinical studies (Sherar *et al.*, 1997a; Lee *et al.*; Ohguri *et al.*, 2018; Kroesen *et al.*, 2019). Thus, while it does not capture all hyperthermic effects, CEM43 has been identified as the an important measure of the treatment efficacy when combining radiation with HT to treat cancer (van Rhoon, 2016). This correlative effect is expected in thermochemotherapy treatments, but not yet validated. Note that all these dosimetric parameters are relevant for treatment efficacy, but many are also used to assess treatment safety, for which $T_{max}$ in healthy tissue is often used (Bakker *et al.*, 2018). Finally, the biological equivalent dose (BED) is another dosimetric parameter currently under investigation to capture the combined effect of radiation and hyperthermia treatments (Kok *et al.*, 2022; Kok *et al.*, 2023).

Challenges and Future Directions

The efficacy of HT combined with RT, CT and/or immunotherapy has been explored across various tumour types. Comparing thermal dosimetry is a challenge, because consensus on HT delivery and reporting of dosimetric parameters is limited leading to substantial heterogeneity of HT treatment protocols and clinical results (Ademaj *et al.*, 2022). For instance, tumour $T_{min}$ was a prognostic factor in a few studies (Cox and Kapp, 1992) and another study showed that $T_{90}$, $T_{50}$, and $T_{10}$ in the target volume were more strongly correlated with cancer response than tumour $T_{min}$ (Leopold *et al.*, 1992). Moreover, a short time interval between HT and RT was shown to significantly predict treatment outcome in retrospective analyses of cervical cancer patients (van Leeuwen *et al.*, 2017). Note that conflicting results have been also reported (Kroesen *et al.*, 2019), which may be attributed to differences in time interval and tumour temperature achieved, and in the patient population included (Crezee *et al.*, 2020). Thermal dose was successfully tested in several clinical trials as predictor of RT + HT tumour response (Jones *et al.*, 2005; Franckena *et al.*, 2009; Bakker *et al.*, 2019; Ademaj *et al.*, 2022).

The thermal dose parameters used to date did not lead to established thresholds for different cancer sites, even though the European Society for Hyperthermic Oncology (ESHO) guidelines recommend superficial HT treatment maintain $T_{50} \geq 41$ °C and $T_{90} \geq 40$ °C (Dobšíček Trefná *et al.*, 2017). Similarly, safety thresholds are yet to be defined. However, several reports indicate that if temperatures reach levels above 44-45 °C, patients can experience pain, formation of oedema, and thermal blistering for local applications and solid organ damage for regional therapies (Linthorst *et al.*, 2015; Longo *et al.*, 2016). Temperatures within 43-44 °C may be safe for short periods of time if sufficient thermometry is provided to maximize tumour monitoring coverage (Bakker *et al.*, 2018).

### A. Probe-based thermometry

Invasive or intraluminal temperature measurements performed directly within the tumour are still regarded by many clinicians to be the gold standard for monitoring HT treatments. Temperature probes used to monitor HT treatment include thermistors, optical fibre probes and thermocouples (Dobšíček Trefná *et al.*, 2017; Schooneveldt *et al.*, 2016). These are inserted into preplaced closed-tip catheters and can be arranged in multi-sensor probes, or cyclically pulled though the catheter, to increase spatial resolution. Thermocouples include a metal component that makes them sensitive to EM fields and requires power to be momentarily stopped (few seconds) to reduce the effects of self-heating (Bakker *et al.*, 2020; De Leeuw *et al.*, 1993). In superficial HT, probes are positioned on the surface and sometimes invasive probes are used to measure temperature at depth. In deep HT, temperature is measured at the surface (for control) and via probes inserted into closed-tip catheters that are placed in the rectum, vagina and bladder. Only occasionally, due to the risk of bleeding and tumour spreading, interstitial catheters are inserted directly in the tumour for direct temperature readings. Note that interstitial HT uses brachytherapy catheters for both placement of MW antennas and temperature probes so there is no need to add extra invasive measurements.

The main challenge with point-measurement approaches is the limited number of monitoring positions, which result in under sampling of both tumour and healthy tissue temperatures (Bakker *et al.*, 2018). More frequent insertion of intratumoral probes, on the other hand, will restrict the number of patients being treated with HT due to potential complications associated with the invasive procedure. Hence, a noninvasive and 3D monitoring approach is thus warranted.

### B. MR thermometry

Noninvasive magnetic resonance (MR) thermometry, or MR thermal imaging (MRTI) is emerging as an alternative to invasive and point measurements (Winter *et al.*, 2016). Most commonly, the shift of the proton resonance frequency (PRF) is used due to its linearity with temperature and independence of tissue type (Rieke and Butts Pauly, 2008a). In this approach, baseline MR images are acquired before treatment to map the PRF. During HT, the PRF is measured at set time-intervals, which is subtracted from the baseline PRF map to monitor the local temperature change in each voxel by the PRF shift. Despite few companies providing MR-guided thermal therapy treatments at ablative temperatures, there is only one device for MR-guided hyperthermia commercially available for deep-regional HT treatments (Curto *et al.*, 2019).

As the PRF shift with temperature is a very small effect, measurement noise is severe and other variables like patient or physiological motion (cardiac or respiratory) create inaccuracies and/or severe artifacts (Feddersen *et al.*, 2020). Advanced algorithms and techniques are employed to correct these inaccuracies and artifacts. For example, while the PRF shift provides temperature change in water-rich tissues, the limited PRF shift in low-water tissues like fat can be used to correct the images for other sources of PRF changes (Baron *et al.*, 2014), e.g., motion effects but also MRI scanner drift (Bing *et al.*, 2016). Still, these techniques are mostly at the research phase (VilasBoas-Ribeiro *et al.*, 2021).

### C. Hyperthermia treatment modelling

Pre-treatment planning by HT modelling involves using simulation tools to optimize the dose distribution before treatment (Paulides *et al.*, 2013). First, imaging slices are used to segment tissue regions, which is often a semi-automatic process (Ribeiro *et al.*, 2018). Second, each region is assigned EM and thermal tissue properties to obtain 3D EM and thermal representations of the patient. Third, EM field simulation is conducted to predict how the energy will be distributed within the tissues for each antenna, which will be used to calculate 3D SAR from multiple antennas in step four. Following this, temperature calculations can be made to estimate the tissue's thermal response to the applied energy. This workflow is illustrated in Figure 1. Pre-treatment planning is generally used for phased-array applicators, for which phase-amplitude fine-tuning of the signals applied to the antennas is carried out to achieve maximum heating, i.e., predicted SAR or temperature, while minimizing the impact on surrounding healthy tissues (Paulides *et al.*, 2021).

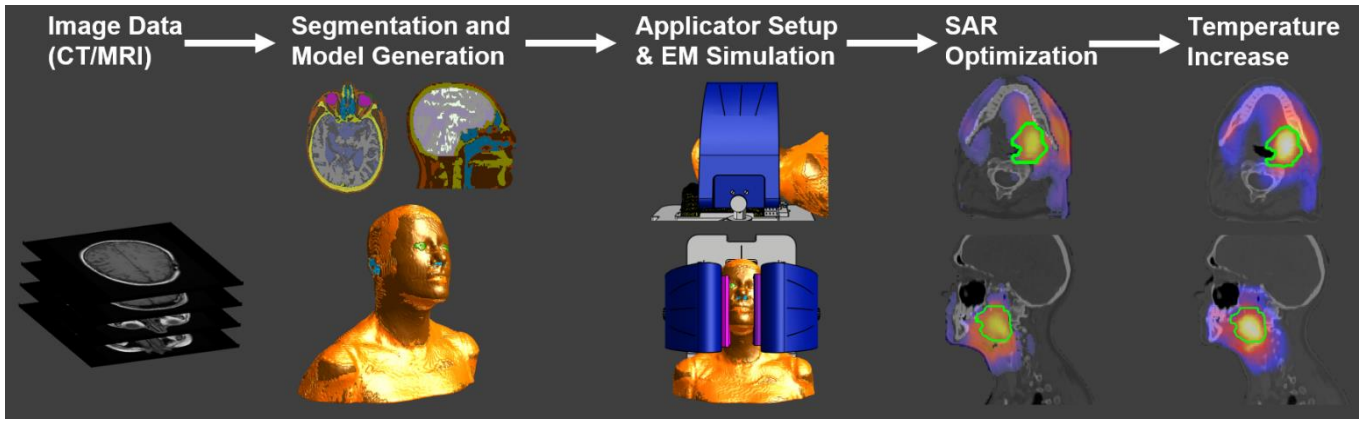

Figure 1. Schematic overview of the hyperthermia treatment planning workflow, illustrated using a head and neck hyperthermia applicator named MRcollar that operates at 434 MHz (Paulides *et al.*, 2018). Adapted with permission from Rodrigues *et al.* (2022).

One of the main challenges in patient HT modelling is the unknown EM properties of the patient tissues, so literature-based values are assigned for all cases. This implies that tissue homogeneity is assumed within normal and tumour tissue, and that there are no heating-induced changes, both of which are oversimplifications. Another challenge is that accurate patient positioning is required for predictive simulations, but recent studies found that patient's position and anatomy during imaging and treatment differ up to 2 cm (VilasBoas-Ribeiro *et al.*, 2023) under MR guidance, and likely even more without it. Moreover, thermal tissue properties are temperature and blood perfusion dependent, and the body counteracts tissue heating using strong, local (immediate) and global (after approximately 15 min for significant heat loads) thermoregulation responses. Consequently, temperature-based models have great uncertainties at this moment. Therefore, although pre-treatment planning is advised in ESHO protocols (Bruggmoser *et al.*, 2011), these uncertainties have limited its use in wide-scale routine clinical practice. Besides temperature, modelling the biological effects of HT combined with other modalities is crucial to understand how tissues respond to different levels of exposure (Kok *et al.*, 2022). However, this type of modelling has even greater uncertainties and only provides indicative values. Finally, the high variability and uncertainty of patient tissue properties lead to a sharp difference in patient dosimetry between HT and radiotherapy: patient-specific thermal dosimetry can currently only be calculated at an adequate accuracy after collecting the temperature data.

**D. Phantom dosimetry**

Phantom dosimetry plays a major role in assuring that HT applicators function correctly, as well as for experimental validation of novel heating and monitoring approaches. Moreover, phantom dosimetry is crucial for comprehensive assessment of the applicator modelling implementation, both at installation and periodically thereafter, in view of HT modelling. To quickly assess the heating symmetry and steering abilities of phased-array applicators, quality assurance procedures use LED or Lamp phantoms with cylindrical or elliptical shape (Aklan *et al.*, 2019; Wust *et al.*, 1994; Schneider and Van Dijk, 1991). These consist of a LED diode or lamp matrix immersed in a saline solution contained in a cylindrical/elliptical plastic shell. The saline is calibrated to match tissue electrical conductivity enabling the LEDs/lamps to respond to the generated EM field. The fast assessments can be effectively carried out qualitatively by using LED/Lamp phantoms or quantitatively either by using the Schottky diode sheet (Van Rhoon *et al.*, 2003) or even the so-called wallpaper paste phantom with multiple thermal mapping probes covering axial and longitudinal measurements in catheters.

An important challenge impeding widespread application of hyperthermia treatment modelling, for both pre-planning and real-time treatment guidance, is the current lack of rigorous device performance verification. The limited commercial solutions to measure the heating delivered by the devices generally fail in terms of the quantitative SAR assessment required to assure simulation accuracy.

### Scientific and Technological Advances Addressing the Challenges

MR thermometry is expected to become a crucial technique in hyperthermia treatment research and possibly also in the clinical routine, as new advances enable it to ensure heating as well as a sound validation of treatment modelling. Hereto, new MR compatible microwave devices are under development that enable more precise focus steering possibilities and can contain integrated coils to reduce the noise of the imaging, leading to better motion correction possibilities (Drizdal *et al.*, 2021; Sumser *et al.*, 2021). In addition, new correction approaches are under development with various (Kok *et al.*, 2022) levels of complexity, addressing challenges such as motion artifacts and improving the reliability of temperature measurements (Nouwens *et al.*, 2022). Additionally, MR enables improved tissue modelling, allowing for more accurate predictions of thermal responses and better treatment modelling. This treatment modelling in turn enables hybrid thermal assessment for filling in the gaps of imaging in both space and time (VilasBoas-Ribeiro *et al.*, 2022). Together, these innovations are set to significantly advance the field of HT.

Adaptive treatment modelling allows for adjustments to be made during treatment based on real-time data like from temperature sensors or patient complaints, i.e., the patient generally senses overheating normal tissue through pain before the heating leads to (severe) side effects. Adaptive HT based on model guidance is currently used routinely in a few specialized centres. Given that relative agreement between model and measurements are already adequate, this approach is already providing clinical benefits (Franckena *et al.*, 2010; Kok *et al.*, 2014). In addition, as patient positioning and posture control is improved and our knowledge on EM and thermal tissue properties expands, robust planning and re-planning approaches can be exploited. Furthermore, the treatment model also enables us to include real-time supportive measurements from sources like E-field sensors or flow sensors that are placed inside the device. This helps to enhance the predictive value of the model, which is expected in turn to translate into improved benefits of adaptive HT.

Generally, it is assumed that the developments in simulation and MRT guided treatment will reinforce each other since MRT forms the tool required to validate treatment planning, patient positioning and device performance in 3D (VilasBoas-Ribeiro *et al.*, 2024). Note that validated treatment planning simulations not only provide the tool for guiding treatment, but also provides the tool to design better applicators, create treatment and quality assurance guidelines, conduct superior training, develop new treatment approaches, and provide more robust thermal dose assessments of patient treatments (Paulides *et al.*, 2021). The latter are crucial to establish correlations between thermal dose and treatment outcome or toxicity per indication and per multi-modality treatment approach.

Concluding Remarks

In summary, hyperthermia treatment continues to evolve through improved dosimetry, real-time monitoring, and adaptive modelling. Integration of MR thermometry and simulation tools is key to enhancing clinical outcomes and treatment safety. Continued research and standardization efforts are critical for establishing robust treatment protocols and expanding clinical adoption.

Acknowledgement

MMP was financially supported by Eurostars project 3868 (SENS-THERM) and project P21-34 (CARES) of the research programme Perspectief, which is financed by the Dutch Research Council (NWO).

*4.2 Radiofrequency and Microwave Ablation*

C. L. Brace[1], M. Cavagnaro[2], P. Prakash[3]

[1] Departments of Radiology and Biomedical Engineering, University of Wisconsin-Madison, USA

[2] Department of Information Engineering, Electronics, and Telecommunications, Sapienza University of Rome, Italy

[3] Department of Biomedical Engineering, The George Washington University, USA

Status

Thermal ablation is a medical treatment where power from an external source is delivered to tissue via an applicator. The applied power is locally absorbed leading to increased tissue temperatures and tissue destruction. Thermal ablation systems are used for various applications in oncology, cardiac electrophysiology, dermatology, gastroenterology, neurosurgery, and others. In this section of the roadmap article, we will restrict our attention to thermal ablation employing RF (~500 kHz) and microwave (~500 MHz – 10 GHz) energy.

The temperature field created during an ablation procedure is a function of the electromagnetic power loss density around the applicator and factors contributing to bioheat transfer, notably: thermal conduction; microvascular blood perfusion and heat transfer due to flow in discrete vessels; active cooling of ablation applicators; and others. Advances in experimental and computational assessment of transient temperature profiles have contributed to the development and advancement of thermal ablation devices and systems, and have potential to inform treatment delivery strategies in a variety of tissue types.

As previously introduced, thermal ablation employing radiofrequency (RFA) and microwave (MWA) energy use two different frequency ranges. This difference translates into a different mechanism of energy release, which in turn leads to different possible outcomes of the two techniques.

During RFA, an electric current is released by an electrode into the tissue to be treated. While propagating into the tissue, the current is dissipated through the Joule effect: the electric power lost by the current is transformed into heat that increases the tissue temperature. At about 500 kHz, the wavelength is much larger than the electrode dimension or the tissue to be treated (e.g. in muscle the wavelength is about 10 m at 500 kHz). Accordingly, the electrode can be modelled as a point source, and the current follows radial paths while moving away from the electrode. This means that the current spreads out on concentric spheres, i.e., the current density amplitude decreases according to a law inversely proportional to the square of the distance from the electrode. Given that the heat is proportional to the dissipated power, which in turn is proportional to the square of the current density, the heat source into the tissue has an amplitude that depends on the distance from the electrode ($r$) as $1/r^4$. The direct consequence of this dependence is that the electric current directly heats (thermally ablates) only a few mm of tissue around the electrode, while the overall RF ablation zone is mainly achieved by thermal conduction, which is a slow phenomenon, easily impeded by blood flow (Huang, 1995).

Still related to the physics of RFA, electric current flows through tissues with higher water and ionic content. However, if the tissue's temperature reaches about 100 °C, water vaporization and tissue desiccation increase the circuit impedance, causing electric power to drop (Chu and Dupuy, 2014). Expandable electrodes, power cycling, and saline infusion are common techniques to help distribute the electrical energy and control heating by RFA devices.

At the frequencies of MWA, an EMF is radiated by an antenna within the MWA applicator. The most commonly used frequencies for MWA in clinical use are 2.45 GHz and 915 MHz, though systems operating at higher frequencies have also been explored (Jones *et al.*, 2012; Curto *et al.*, 2015; Bottiglieri *et al.*, 2022). In contrast to the electric current in RFA, the EMF propagates in lossy as well as lossless materials, so that no theoretical limitation exists on the maximum reachable temperature. While propagating, the EMF dissipates due to losses into the tissue; the lost power is transformed into heat. In MWA, the temperature increase is directly proportional to the electromagnetic power dissipation, roughly decreasing according to a law inversely proportional to the square of the distance from the antenna. Accordingly, the thermally ablated area is mainly due to the electromagnetic dissipation, a phenomenon much faster than thermal conduction (Brace, 2009).

B. Dosimetric Parameters and Quantities of Interest

The Appendix presents the pertinent physics and equations for characterizing time-averaged EMFs, power absorption, and bioheat transfer during RFA and MWA. The biological effects of thermal therapy are a function of the time-temperature history during heating (Pearce, 2013). In vitro studies of thermal exposure to cells in culture to identify the kinetics of thermal injury have demonstrated that the time taken to observe a bioeffect (e.g. cell death) decreases with increasing temperature. Cell death occurs rapidly at temperatures above ~55 °C, and as such threshold temperatures in the range ~50-60 °C are widely used to approximate the extent of cell death. Thus, the primary quantity of interest for dosimetry of ablation system is the transient temperature profile, $T(\underline{r},t)$ [°C] in tissue. Tissue temperature can be measured directly with a variety of invasive temperature sensors (e.g. thermocouples, thermistors, fibreoptic sensors), or estimated via medical imaging techniques (Zaltieri *et al.*, 2021).

The time-averaged electromagnetic power loss density profile $Q(\underline{r})$ [W/m³] is another quantity of interest as it represents the locally absorbed power in tissue and provides insight into the region of tissue where the applied power directly affects tissue.

$$Q(\underline{r}) = \rho(\underline{r}) SAR(\underline{r})$$

As previously reported, the Specific Absorption Rate ($SAR(\underline{r})$) represents the power absorbed in tissue per unit mass, and it is typically used to represent the electromagnetic heat source into the bio-heat equation. It can be evaluated from measurements of the electric field once the conductivity and density are known, or from the transient temperature profile, when measured using very short time durations in phantoms (Deshazer *et al.*, 2017).

One of the main difficulties in using these equations is the high dose rate during RFA and MWA. In contrast to HT, thermal ablation is of relatively short duration, typically on the order of ~5-15 min for oncologic applications, and often less for other applications where the treatment zone is small. In this time temperatures can exceed 100 °C, or even much higher during MWA. Therefore, the dosimetry problem is largely focused on achieving an ablative threshold exceeding ~50 °C throughout the targeted tissue, while limiting temperature in non-targeted tissue beneath safety thresholds.

Such dynamic heating also presents a dosimetry problem in the assumptions about electromagnetic and thermal parameters. All tissue properties depend on the tissue temperature, so the electromagnetic and thermal problems become interrelated. However, knowledge about the behaviour of tissues at ablative temperatures (up to about 120 °C) is still quite scattered. Thermal dosimetry has been important for guiding studies aimed at investigating how tissue physical properties and state (e.g. dielectric properties, thermal properties) vary as function of temperature and rate of heating (Deás Yero *et al.*, 2018; Ji and Brace, 2011; Lopresto *et al.*, 2019; Etoz and Brace, 2019; Lopresto *et al.*, 2012; Vidjak *et al.*, 2024). Improved accounting of these temperature dependencies may enable more accurate dosimetry predictions in future studies.

## C. Routine clinical use

In routine clinical use, careful thermal and/or electromagnetic dosimetry are rarely used intra-procedurally. During RF ablation procedures, electrode tip temperature is often monitored with temperature sensors incorporated into the applicator, though these measurements are not directly indicative of extent of the ablation zone. For tumour ablation procedures, the outcome is typically assessed with contrast imaging to assess non-enhancing regions, indicative of tissue regions devoid of perfusion following the ablation (Lin *et al.*, 2024). Recently, image analysis tools have been developed and are in clinical use for assessing the extent of the ablation zone relative to the tumour boundary prior to ablation, to facilitate estimation of treatment margins (Laimer *et al.*, 2025). Several clinical systems accommodate the use of auxiliary temperature sensors that clinicians may choose to position at regions at risk to monitor for thermal damage or insufficient treatment.

Volumetric thermometry with MRI during RFA and MWA has been demonstrated in the clinical setting and has potential to provide transient temperature profiles in multiple planes (Lepetit-Coiffé *et al.*, 2010; Öcal *et al.*, 2024). However, this technique is not routinely used due to technical complexity of integrating ablation systems within the MRI environment, the considerable added cost of conducting clinical procedures in the MRI suite, and the potential for tissue motion and water content changes to corrupt the thermographic data. This remains an area of ongoing research and development.

Careful thermal dosimetry has played a key role in guiding the development of ablation applicators and energy delivery strategies. For example, several studies over the last three decades measured temperature at select distances from ablation applicators to develop insight into how various device and/or energy delivery parameters affected the overall treatment zone (Yang *et al.*, 2007; Cavagnaro *et al.*, 2011; Andreano and Brace, 2013; Curto *et al.*, 2015; Namakshenas *et al.*, 2024). Thermal dosimetry has also been important for understanding how the physical quantity that ablation procedures ultimately modify (i.e. temperature) relate to quantities measured/observed in the clinical setting (e.g. imaging appearance of tissue exposed to ablation with/without contrast) (Strigari *et al.*, 2019). In each case, limitations in spatial or temporal resolution, or a limited range of tissue types and/or physiological variables such as perfusion, impacted their extrapolation into broader clinical utility.

Challenges/opportunities

Several challenges persist in verifying adequate thermal coverage of the target zones across clinical indications.

While most device manufacturers provide guidelines for expected ablation size compared to power and delivery time, they are often characterized in *ex vivo*, unperfused, animal tissue (Hoffmann *et al.*, 2013). Ex vivo ablations are not necessarily representative of ablation zones achieved in the clinical setting due to the influence of blood perfusion heat sinks, organ boundaries, and heterogeneity in tissue physical characteristics as a function of disease state (Amabile *et al.*, 2017). Accordingly, there remains a gap between the expected and realized treatment effects for similar equipment settings.

Computational approaches for estimating tissue temperature profiles during ablation are under investigation for potential use in planning treatments and/or assessing treatment outcomes. Physics-based models that employ numerical approaches to solve the differential equations pertinent to electromagnetic absorption and bioheat transfer have been applied towards predicting extents of the treatment zone, as estimated from computed transient temperature profiles. Recent efforts in the field have integrated these models with pre- and peri-procedural imaging data to inform model geometry and facilitate model predictions of treatment zone overlaid with imaging.

A few clinical studies evaluating the technical feasibility and performance of these computational approaches have been reported (Moche *et al.*, 2020; Hoffer *et al.*, 2024).

Real-time or immediate post-treatment feedback from thermal sensors or medical imaging can help ensure that the proper thermal dose was used for a full treatment effect, but this approach may not be practical for all clinical situations. Invasive temperature probes can provide continuous feedback with limited spatial resolution. In applications where the ablation zone is relatively small ($\leq 1$ cm maximum dimension), such limited spatial resolution may be sufficient; however, for large ablations ($\geq 3$ cm maximum dimension) such pointwise data cannot give reliable feedback about the entire treatment volume (Zaltieri *et al.*, 2021). Local heterogeneities in tissue perfusion, structure or other properties can lead to irregular ablation shapes that deviate from expectations. Substantial variation is expected between patients as well (Mathy *et al.*, 2024). Imaging-based dosimetry can fill some of the gaps to provide more full-volume thermal data, but at the cost of reduced temporal resolution and generally greater noise figures than direct thermal profiling. Recent studies have explored alternative solutions to determine noninvasively temperature distributions in real time, such as the use of microwave imaging (Wang *et al.*, 2022b) or analysing tissue dielectric property changes from the antenna applicator (Vidjak *et al.*, 2023; Wang and Brace, 2012).

Concluding Remarks

The thermal profile induced in tissue during thermal ablation can vary considerably across procedures using the same applied energy settings, potentially contributing to the variability in treatment outcomes. Currently, there are few practical means for thermal dosimetry throughout targeted tissue volumes in routine clinical use. Computational and experimental approaches for estimating volumetric thermal profiles are the subject of active research investigation and have potential to contribute to improved outcomes of thermal ablation procedures.

Appendix:

## A. Relevant physics/mathematical problem

The mathematical problem in both RFA and MWA is stated by Maxwell's equations and the so-called bio-heat equation. Maxwell's equations can be solved either in the time or in the frequency domain (Cheng, 2014):

$$\nabla \times \underline{E}(\underline{r}, \omega) = -j\omega\mu\underline{H}(\underline{r}, \omega)$$

$$\nabla \times \underline{H}(\underline{r}, \omega) = j\omega\varepsilon_0\varepsilon(\omega)\underline{E}(\underline{r}, \omega) + \sigma_s\underline{E}(\underline{r}, \omega)$$

where $\underline{r}$ represents a generic position vector in space, $\omega$ is the angular frequency, $\underline{E}$ the electric field vector [V/m], $\underline{H}$ the magnetic field vector [A/m], $\mu$ [H/m] the magnetic permeability that can be considered equal to that of the vacuum $\mu_0$ [H/m], $\varepsilon_0$ the permittivity of vacuum [F/m], $\varepsilon(\omega)$ the complex relative permittivity, and $\sigma_s$ [S/m] the static conductivity.

When RFA is concerned, quasi-static conditions can be applied, so that the electromagnetic problem is solved through the Laplace equation (Cheng, 2014).

$$\nabla \cdot \left(\sigma_s(\underline{r})\nabla V\right) = 0$$

where $V$ represents the potential linked to the electric field through

$$\underline{E}(\underline{r}, t) = -\nabla V(\underline{r}, t)$$

Boundary conditions are needed to solve Maxwell's equations, as well as initial conditions when the time domain formulation is considered.

The temperature increase induced by the absorption of the EMF can be determined by solving the bio-heat equation (Pennes, 1948)

$$C(\underline{r})\rho(\underline{r})\frac{\partial T(\underline{r}, t)}{\partial t} = \nabla \cdot \left(K(\underline{r})\nabla T(\underline{r}, t)\right) + A_0(\underline{r}) + \rho(\underline{r}) \cdot SAR(\underline{r}) - B_0(\underline{r})\left(T(\underline{r}, t) - T_b\right)$$

with the corresponding initial

$$\nabla \cdot (K(r)\nabla T) + A_0(r) - B_0(r)(T - T_b) = 0$$

and boundary conditions

$$-K\left(\frac{\partial T}{\partial n}\right)_s = H(T_s - T_a)$$

In the previous equations, $T$ is the temperature [°C], $C$ the specific heat [J/(kg °C)], $\rho$ the density of the tissue [kg/m$^3$], $K$ the thermal conductivity [J/(s m °C)], $A_0$ the volumetric heat generation linked to metabolic processes [J/(s m$^3$)], $B_0$ is a quantity proportional to blood perfusion [J/(s m$^3$ °C)], $T_b$ is the blood temperature, $H$ the convective coefficient [J/(s m$^2$ °C)], $T_s$ the temperature at the surface of the tissue, $T_a$ the temperature of the surrounding material, and $SAR$ represents the Specific Absorption Rate, evaluated as

$$SAR = \frac{\sigma |\underline{E}|^2}{2\rho}$$

where $\sigma$ represents the equivalent conductivity including both the static conductivity and the dielectric losses (the imaginary part of the relative permittivity).

# 5. Radiofrequency-based Modalities for Diagnosis and Sensing

## 5.1. Radiofrequency Dosimetry in Ultra-High Field Parallel Transmit MRI

Desmond T. B. Yeo

GE HealthCare, Technology and Innovation Center, Niskayuna, NY, USA

*Status*

Due to its superior soft tissue contrast, diverse functional and structural information content, and absence of ionizing radiation, magnetic resonance imaging (MRI) has had significant impact in the management of multiple diseases and disorders. MRI requires the application of i) a strong static magnetic field ($B_0$) to create non-zero net magnetization vectors, ii) circularly polarized radiofrequency (RF) transmit magnetic fields ($B_1^+$) at the Larmor frequency ($f_0$) to nutate spins and (iii) gradient fields for spatial localization. While high $B_0$ fields may induce reversible physiological effects (Schenck, 2000), and fast-switching gradient fields may cause peripheral nerve stimulation (Mansfield and Harvey, 1993), thermal risks in an MRI scan arise primarily from RF-induced concomitant and conservative electric fields (E-fields). The increasing proliferation of very high $B_0$ field MRI systems ($\geq 5T$) with multichannel RF transmit capabilities warrants increased rigor in global/ local RF heating assessment.

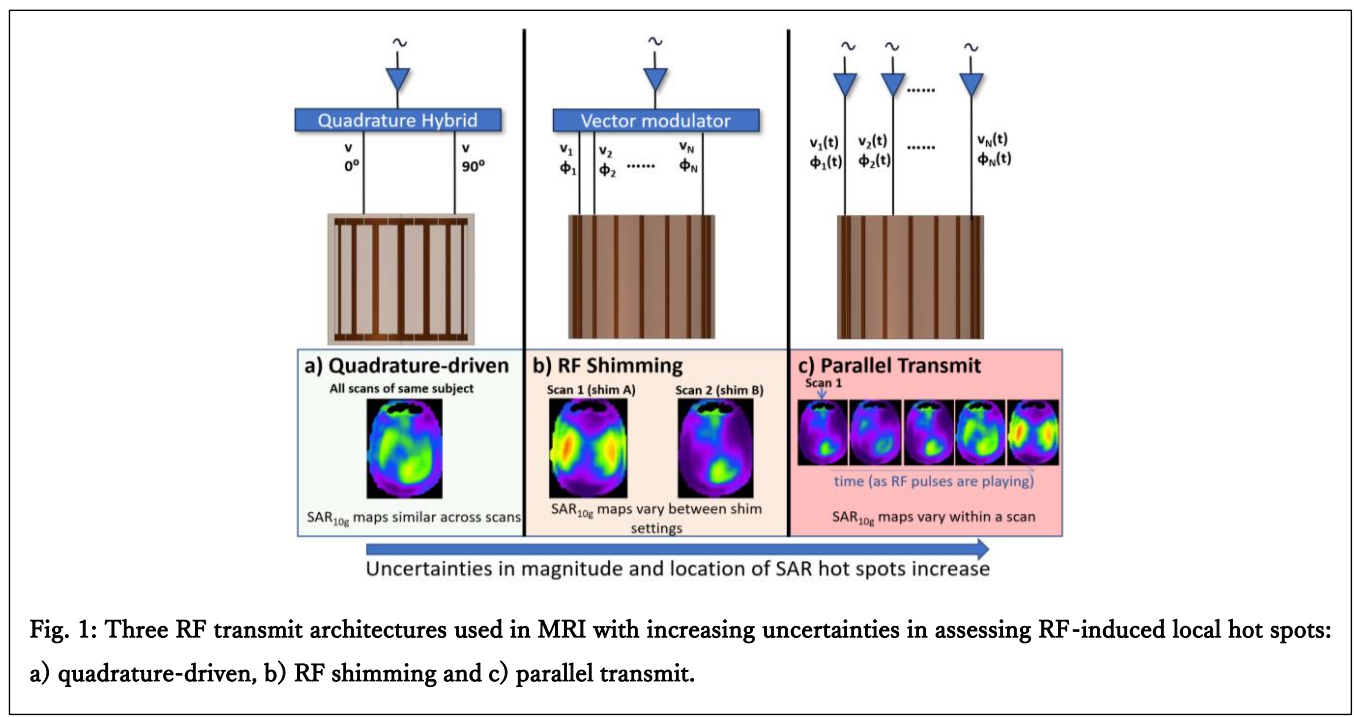

Fig. 1: Three RF transmit architectures used in MRI with increasing uncertainties in assessing RF-induced local hot spots: a) quadrature-driven, b) RF shimming and c) parallel transmit.

As $B_0$ increases, MRI signal-to-noise ratio increases, which improves image quality and/or spatial resolution. Since $f_0$ scales linearly with $B_0$, higher-frequency RF transmit magnetic fields are needed to excite spins as $B_0$ increases, e.g. $f_0$=298.0 MHz at 7T. As $f_0$ increases, the wavelengths of electromagnetic (EM) transmit fields decrease and approach the dimensions of body structures, which induce inhomogeneous $B_1^+$ fields that lead to image shading (Röschmann, 1987; Bernstein *et al.*, 2006) that can confound diagnosis. To address this, multichannel RF transmit architectures and imaging techniques have been proposed, i.e., RF shimming (Hoult, 2000; Ibrahim *et al.*, 2000) and parallel transmit (PTx) (Katscher *et al.*, 2003; Zhu, 2004).

The goal of RF shimming (Fig. 2b) is to create a homogeneous $B_1^+$ map by linearly combining $B_1^+$ fields generated by N transmit elements. Because RF pulse envelopes in all channels are similar in RF shimming, the composite $B_1^+$ and E-field/ local specific absorption rate (SAR) distributions do not change appreciably during RF excitation, and throughout an exam for a given set of complex weights. Conversely, in PTx MRI (Fig. 2c), which can also be used to accelerate spatially selective excitation (Katscher *et al.*, 2003), each transmit channel's signal is independent of other channels. Instead of a static, amplitude-weighted $B_1^+$ distribution, PTx RF pulses produce time-varying $B_1^+$ distributions, which induces time-varying concomitant E-field/ local SAR distributions. While the additional degrees of freedom afforded by PTx MRI can potentially reduce local SAR, the absence of accurate personalized information can increase risks of tissue heating instead (Zhu, 2004).

In 1.5T and 3T MRI scanners that utilize quadrature-driven RF transmit volume coils (Fig. 2a), RF power deposition in the exposed mass is *predicted* prior to a scan using knowledge of the RF pulse envelope and patient weight. During the scan, global average SAR in the exposed mass is then *measured* with power monitors in the RF transmit chain, typically in tandem with pre-characterized information that account for coil loss, exposed mass, etc. The **predicted** and **measured** power deposition rates are time-averaged (6 minutes and 10 seconds) and assessed against whole body and/or partial body global average SAR limits stipulated in IEC 60601-2-33 (2022). In the same way, PTx coils require SAR prediction and monitoring mechanisms to ensure patient safety. However, depending on how a PTx coil is used, both global and **local** SAR limits may be pertinent (IEC 60601-2-33, 2022; Fiedler *et al.*, 2025). Compared to quadrature-driven coils, the higher uncertainty and potential severity of local SAR hot spots in PTx MRI present a more challenging RF safety problem that often leads to the implementation of high local SAR safety margins at the expense of imaging performance, e.g., worst-case local SAR predictions (Neufeld *et al.*, 2011). Conversely, a framework that consistently under-predicts local and global SAR would be unacceptable for patient safety. In PTx MRI, it is desirable to develop frameworks that minimize over-prediction of SAR, with safeguards that prevent under-prediction. In the long term, since thermal dose and thermal damage thresholds are more predictive of heat-induced tissue damage (van Rhoon *et al.*, 2013; Yarmolenko *et al.*, 2011), frameworks that measure subject-specific local temperature rise and thermal dose over time can increase confidence when assessing RF heating risks in ultra-high field (UHF) PTx systems.

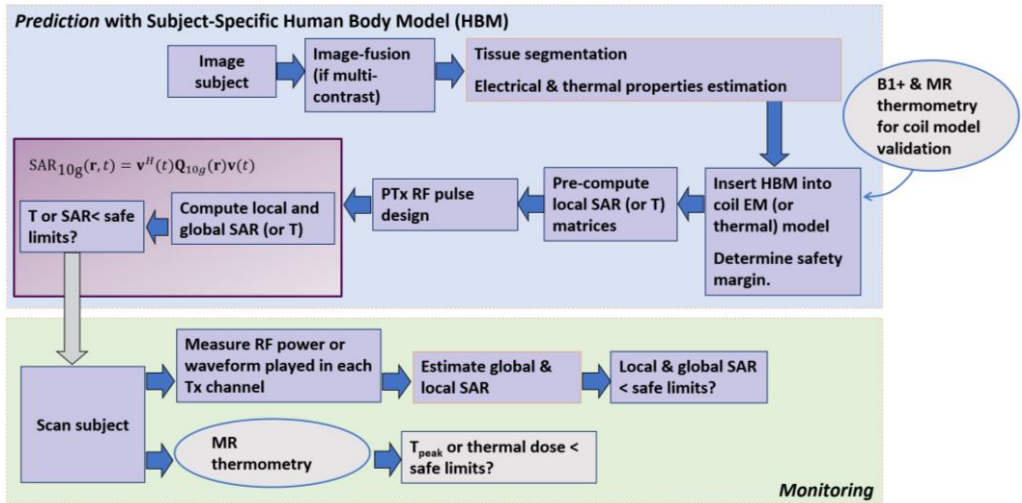

Fig. 2: Framework for personalized RF dosimetry in PTx MRI.

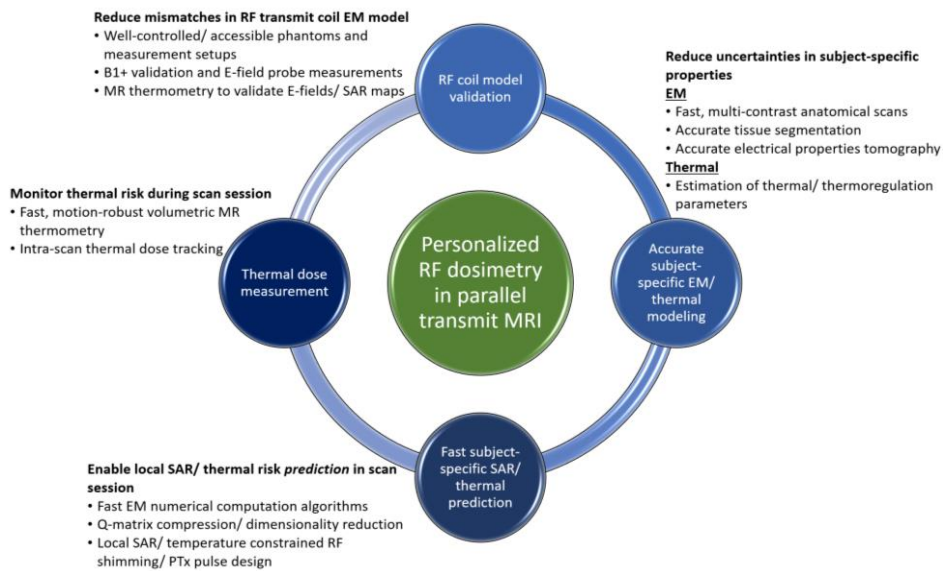

Fig. 3: Challenges and advances needed to enable efficient, personalized RF dosimetry in PTx MRI.

Current and future challenges.

To widen adoption of UHF PTx MRI, it is imperative to increase confidence in the validity of SAR/ thermal risk assessment methodologies. Fig. 3 depicts a personalized local SAR/ thermal dosimetry workflow for PTx MRI that incorporates modules that have been studied (Homann *et al.*, 2011; Graesslin *et al.*, 2012; Boulant *et al.*, 2016). The challenges in realizing this workflow can be categorized as follows (Fig. 4): **i)** accurate PTx coil array EM model validation (particularly for E-fields), **ii)** accurate personalized EM/ thermal modelling to minimize prediction errors**, iii)** real-time personalized local SAR prediction/ measurement and **iv)** real-time thermal dose measurement.

The validation of EM coil models in inhomogeneous tissue conditions is limited by the lack of direct high-resolution E-field measurements, often necessitating indirect approaches such as phantom-based MR thermometry (MRT) (Ishihara *et al.*, 1995; Poorter *et al.*, 1995; Rieke and Butts Pauly, 2008b).

In UHF PTx, a safety margin derived from EM modelling with generic human body models (HBM) is commonly applied to over-predict *local* SAR (Neufeld *et al.*, 2011). This is because it is very challenging to accurately replicate the salient input parameters for EM simulations pertaining to the subject, RF transmit coil, landmark position, etc. To create a personalized HBM, whole or partial body MRI datasets with appropriate contrast types need to be acquired efficiently (likely on non-UHF scanners), followed by accurate and fast tissue segmentation, which presents workflow and computational challenges. MR-based electrical property tomography (EPT) can potentially provide personalized estimates of electrical properties but improvements in robustness and computational efficiency are needed (Katscher and van den Berg, 2017; Leijsen *et al.*, 2021). Thereafter, the HBM can be inserted into a validated PTx coil EM model. Fast EM solvers are needed for online computation of EM fields to generate local SAR/ Q matrices (Bardati *et al.*, 1995). These matrices can be applied to PTx RF pulse design algorithms with local/ global SAR constraints (Graesslin *et al.*, 2012; Boulant *et al.*, 2016). Techniques that compress the set of Q matrices can significantly reduce computation time in local SAR-constrained RF pulse design but higher levels of compression are often accompanied by greater over-prediction of local SAR (Eichfelder and Gebhardt, 2011). Solving Pennes' bioheat equation across transmit channel combinations yields a Q matrix-equivalent for temperature rise (Das *et al.*, 1999), useful for predicting heating under different PTx waveforms. However, accurate in vivo temperature prediction requires subject-specific thermal property maps (e.g., perfusion, conductivity) and modelling of nonlinear thermoregulation, both of which remain challenging (Murbach *et al.*, 2014).

In transitioning from pre-scan local SAR *prediction* to intra-scan local SAR *measurement*, the problem becomes more challenging as MRI cannot directly measure E-field distributions. Even if personalized local SAR/ temperature predictions are accurate, they do not account for in-scan events like transmit channel failures, which can produce E-field distributions that differ from predictions. A local SAR monitoring approach might consist of channel-specific sniffer coils that partially measure each channel's transmit field information, which is then applied to precomputed, simulated E-fields derived from individually excited channels that irradiate a personalized HBM. This indirect monitoring of local SAR will need to account for uncertainties in the field measurement process. Alternatively, volumetric MRT techniques (Ishihara *et al.*, 1995; Poorter *et al.*, 1995; Rieke and Butts Pauly, 2008b) can be applied to measure temperature and thermal dose maps during a scan, which may obviate the reliance on personalized HBMs or sniffer coils in the monitoring process. However, it is challenging to implement fast, motion robust, volumetric MRT techniques that are accurate and precise enough to detect tissue heating across a wide field of view (FOV) in real-time. Realistic thermal dose thresholds and motion-robust volumetric MRT remain challenging for online monitoring.

Advances in science and technology to meet challenges.

Advances that lead to accurate and precise estimates of subject-specific parameters for EM/ thermal modelling can reduce mismatches between actual scans and simulation results. Fast segmentation of multi-contrast MRI datasets, combined with atlas-based nonlinear registration, can efficiently generate personalized HBMs for local SAR prediction (Brink *et al.*, 2022; Cabezas *et al.*, 2011). In addition to enabling personalized dosimetry, these segmentation approaches can be used to generate large numbers of realistic, non-subject-specific HBMs with diverse characteristics that impact local/ global SAR risks. Data-driven algorithms that learn salient features from tissue-segmented imaging datasets may conceivably be developed to generate synthetic HBMs to augment existing repositories. If widely accessible and thoroughly validated by experts, the repositories can help reduce uncertainties when determining safety margins for a broad population.

Direct MR EPT methods that estimate electrical property maps from measured transmit/ receive fields are computationally less expensive than forward methods (Leijsen *et al.*, 2021). However, if forward methods, which require tissue-segmented data and EM numerical computation, can be accelerated sufficiently, they can potentially produce more robust estimates of electrical property maps. Another approach, image-based EPT, reconstructs electrical properties directly from MR images, which obviates the need for transmit/ receive fields (Leijsen *et al.*, 2021), i.e., reduces scan time and increases SNR. Emerging data-driven EPT approaches have also shown promise in estimating electrical property maps that are more robust to noise (Leijsen *et al.*, 2022; Mandija *et al.*, 2019).

The use of accurate personalized HBMs and electrical properties within a carefully validated PTx coil model will increase confidence in the validity of output from modules in Fig. 3. To minimize the odds of under-predicting local SAR, formulations like the virtual observation points that compress Q matrices (derived from personalized HBMs) can provide upper limits for fast local SAR estimates that can be adjusted to trade off the degree of compression and the level of local SAR over-prediction allowed (Eichfelder and Gebhardt, 2011). Alternatively, an efficient PTx RF pulse optimization algorithm that can operate on uncompressed Q matrices within a practical timeframe can be employed to minimize over-prediction and obviate the compression process (Pendse *et al.*, 2019). It has also been shown that fast local SAR prediction with low over-prediction errors is feasible with supervised deep learning networks trained with $B_1^+$ and local SAR maps (Meliadò *et al.*, 2020; Gokyar *et al.*, 2023).

Volumetric thermal dose measurements require fast, motion-robust MRT techniques. With proton frequency resonance MRT (Poorter *et al.*, 1995), motion-induced $B_0$ perturbations can create non-temperature-dependent phase accruals that result in MRT errors. These increase in severity as $B_0$ increases. A recent study that utilized motion-compensated MRT demonstrated the feasibility of sub-degree accuracy when monitoring RF-induced heating in the brain at 7T (Le Ster *et al.*, 2022). Other motion robust MRT approaches that are applied to thermotherapy applications can also be assessed for online thermal dosimetry (Kim *et al.*, 2024). To overcome the FOV limitations in MRT for RF thermal dose assessment, it is conceivable that carefully calibrated, hybrid measurement-simulation frameworks that combine local thermal dose measurements and EM/ thermal simulation results may ameliorate the shortcomings of both approaches.

Concluding remarks

The assessment of RF heating risks in UHF PTx MRI is confounded by multiple sources of uncertainties. More accessible, standardized phantoms for EM/ thermal validation, fast/ accurate tissue segmentation and EPT techniques, and fast EM/ thermal computation solvers are important drivers that can reduce these uncertainties. When integrated within an efficient processing pipeline, the workflow should be rigorously optimized for online personalized local RF dosimetry. In the long term, accurate MRT techniques that directly measure temperature rise/ thermal dose will provide the most direct risk assessment of RF heat-induced tissue damage. Advancement in these areas will facilitate safe UHF PTx MRI scans, and hopefully, deliver benefit to more patients.

## 5.2. Millimeter Wave Sensing

Takuya Sakamoto[1] and Francesco Fioranelli[2]

1: Department of Electrical Engineering, Kyoto University, Japan

2: Department of Microelectronics, Delft University of Technology, the Netherlands

Status

The use of microwaves and millimetre waves for human body measurements has attracted significant attention, with many research groups actively investigating radar-based, noncontact human sensing (Fioranelli *et al.*, 2019; Gurbuz and Amin, 2019; Le Kernec *et al.*, 2019; Paterniani *et al.*, 2023). These technologies offer important advantages over more conventional contact sensors, such as wearable devices, by reducing patient discomfort and restrictions. Moreover, fewer privacy concerns are raised than those observed in widely used camera-based systems. Radar systems for human measurements can address not only physical quantities, such as distance, angle, position, and velocity, which are the main targets of classical radar systems, but also human-specific shapes and movements, including posture, limb motion, gesture, and activity (Seyfioğlu *et al.*, 2018; Chen *et al.*, 2021; Gorji *et al.*, 2021; Zhao *et al.*, 2022; Zhu *et al.*, 2022; Kruse *et al.*, 2023). These capabilities open promising applications in fields such as healthcare and entertainment, with specific use cases including in-vehicle or outdoor person presence detection, activity and gesture recognition, fall detection, security screening, and gait analysis (Wang *et al.*, 2014; Seifert *et al.*, 2019; Wang *et al.*, 2019; Abedi *et al.*, 2022; Abedi *et al.*, 2023; Gharamohammadi *et al.*, 2023; He *et al.*, 2024).

Radar systems can also detect small-scale body movements associated with physiological processes, including respiration, heartbeats, and pulse waves (Hong *et al.*, 2019; Piriyajitakonkij *et al.*, 2020; Oyamada *et al.*, 2021; Rong *et al.*, 2021; Uddin *et al.*, 2023; Chowdhury *et al.*, 2024). This has great potential for diagnostic applications, such as respiratory-rate and heart-rate monitoring, heart-rate-variability analysis, pulse-wave-velocity measurement, autonomic nervous system monitoring, and even emotional-state assessment (Sakamoto *et al.*, 2015; Sakamoto, 2020; Zhang *et al.*, 2021; Yin *et al.*, 2022; Zhang *et al.*, 2023; Koda *et al.*, 2024). Such technologies have been applied to medical diagnostics (e.g., irregular breathing, cardiac arrhythmias, and respiratory infections) and health-related monitoring and event detection (e.g., sleep apnea, sleep stages, autonomic nervous system activity, animal monitoring, and sudden changes in health status, including those related to sudden infant death syndrome) (Koda *et al.*, 2024; Sakamoto *et al.*, 2024).

Current and Future Challenges

Radar systems based on wideband signals achieve high range resolution, enabling precise measurement of the distance between objects in the field of view and the radar system. Radar systems with two-dimensional (2D) array antennas achieve high angular resolution, enabling the measurement of the azimuth and elevation angles of objects. Doppler radar systems can estimate the line-of-sight velocity of an object *via* frequency shifts. Therefore, radar data are typically represented as complex-valued functions in five dimensions: time, range, azimuth, elevation, and Doppler frequency.

Although Doppler-time spectrograms are widely used in activity recognition because of their ability to capture velocity changes, they often neglect a wealth of information from other dimensions. For physiological sensing, phase-based analysis is often used by extracting the phase of a signal over time and converting it into displacement as a function of time. This enables the detection of subtle motions, such as those caused by respiration and heartbeats, which are often smaller than deliberate body movements. However, both approaches underutilize the full multidimensional potential of radar data, particularly from modern 2D multiple-input and multiple-output radars.

Another major challenge lies in the diversity of data characteristics driven by individual differences in patients/users and varying measurement conditions. Most studies have been conducted in controlled environments, instructing participants to perform specific activities or remain stationary. However, in real-world scenarios, activities are continuous and transitions seamless, and multiple actions often overlap. Stationary individuals exhibit unconscious movements influenced by factors such as age, physical condition, and cultural background. These individual differences can be exploited for identification and authentication (Alrawili *et al.*, 2024; Kobayashi *et al.*, 2024; Peng *et al.*, 2024); however, they complicate radar-based human sensing and demand robust signal processing techniques. For example, body displacement waveforms from physiological signals vary depending on which part of the human body is measured; respiratory displacements at the chest and abdomen are different, as do heartbeat displacements at the head and legs. In the radar measurements of a person, both the position and posture of the target person and the radar installation position affect the measurement accuracy (Li *et al.*, 2006; Wang *et al.*, 2013; Nahar *et al.*, 2018; Koshisaka and Sakamoto, 2024). Therefore, an important future challenge is the development of a technology that can handle data diversity depending on person- and environment-specific factors while combining multiple radar echoes to achieve accurate measurement of physiological signals.

Another challenge is the integration of radar-based sensing with existing systems and services in medicine and healthcare. Sensor fusion (i.e., combining radar with other sensors) can enhance system performance by leveraging the distinct advantages of different modalities (Zhang *et al.*, 2021). Networks of multiple radar sensors, whether identical systems are positioned to observe a target from various angles or diverse systems with varying frequencies and modulation techniques, can further improve the performance through synergistic effects. The use of radar networks is also crucial for measuring the physiological signals and activities of multiple individuals simultaneously (Li *et al.*, 2020; Singh *et al.*, 2020; Iwata *et al.*, 2021; Koda *et al.*, 2021). This is because a single radar alone cannot constantly capture all echoes from multiple individuals due to blocking and shadowing. These integrations are expected to address the current limitations and enable practical applications of radar-based sensing technologies.

Finally, machine learning is expected to play an important role in addressing these challenges (Nocera *et al.*, 2024). Through pattern recognition and feature extraction, diverse and dynamic radar data can be processed and interpreted (Gurbuz and Amin, 2019). However, the most advanced techniques in the field of deep learning require large, diverse, and representative datasets, which are often unavailable in radar applications. Therefore, publicly available databases and effective data-labeling approaches are crucial for the widespread adoption of machine learning in this field (Gusland *et al.*, 2021; Fioranelli *et al.*, 2022; Yang *et al.*, 2023).

Advances in Science and Technology to Meet the Challenges

Radar-based measurements of human activities and physiological signals have benefited from recent advancements in radar system design and processing capabilities. Some recent advances are summarized as follows:

## 1. Exploiting High-Dimensional Radar Data with Machine Learning

To address the challenge of the underutilization of high dimensionality in radar signals, recent advances in radar signal processing techniques and applied machine learning provide a promising solution to improve the performance of radar-based human monitoring by simultaneously exploiting information in multiple domains (i.e., range, velocity, azimuth and elevation angles, and time). Advanced algorithms, particularly deep learning methods, can automatically extract relevant features from large-scale datasets. These approaches are adept at handling multidimensional data inputs, thereby enabling the simultaneous analysis of temporal, spatial, and velocity-related data. By leveraging large amounts of data, machine learning can discover complex patterns and correlations that traditional signal processing may overlook. In this context, the development of well-curated open-access radar datasets can support the training and validation of machine learning models, thus making radar-based sensing more scalable.

## 2. Addressing Data Diversity and Environmental Variability

Real-world radar measurements vary widely due to individual differences and fluctuating environmental conditions. Although machine learning can adapt to diverse training data, advanced signal processing methods, such as adaptive filtering, multidimensional transforms, and other non-machine-learning-based signal separation methods (e.g., independent vector analysis, nonnegative matrix factorization, sparse-representation-based optimization, robust principal component analysis, and Bayesian inference-based methods) are essential for extracting subtle physiological features. Moreover, integrating multiple radar systems follows the same principle as multi-camera setups in computer vision, combining multiple viewpoints to compensate for occlusions, poor angular resolution, and unfavourable aspect angles, thereby enhancing the overall measurement accuracy. These combined strategies enable robust and practical radar-based sensing applications.

## 3. Integrating Radar with Multimodal Sensing Systems

The integration of radar-based sensing with other systems in medicine and healthcare is another frontier of technological advancement. Ongoing research aims to integrate radar-based health monitoring with telemedicine and smart healthcare systems to enable continuous remote physiological sensing. These advancements are expected to increase the clinical viability of radar-based sensing technologies, making them suitable for real-world deployment in hospitals, homes, and public spaces. Sensor fusion techniques—where radar data are combined with information from other modalities, such as optical, acoustic, or inertial sensors—can help develop synergistic systems that overcome the limitations of any single sensor. Moreover, radar sensor networks deployed at multiple locations/angles or using different frequencies and modulation techniques can improve performance by mitigating issues, such as signal blockage and occlusions, particularly when monitoring multiple individuals with arbitrary positions and postures.

Concluding Remarks

Radar-based human sensing technologies offer transformative potential in healthcare, diagnostics, and entertainment by enabling noncontact measurements with minimal privacy concerns. These systems measure physical and physiological parameters and support applications such as activity recognition, fall detection, and heart-rate monitoring. However, current methods simplify radar data, thereby neglecting the full multidimensional potential of radar-based sensing. Addressing this issue requires advanced signal processing to achieve greater accuracy and robustness. Real-world scenarios add complexity because of continuous overlapping activities and individual differences. Machine learning holds promise for overcoming these challenges by enabling advanced pattern recognition and feature extraction; however, progress has been hindered by a lack of large-scale, labeled, and representative datasets. Therefore, the development of public databases and the exploration of unsupervised and transfer learning methods are necessary. Integrating radars with complementary sensors or using multiple radar systems offers opportunities to enhance the radar performance. With continued innovation, radar sensing is expected to play a key role in the creation of smarter and more connected solutions to meet societal needs.

## 5.3. Terahertz Sensing

Zachary Taylor[1], Emma Pickwell-MacPherson[2] and Maya Mizuno[3]

1 Department of Electronics and Nanoengineering, Aalto University, Finland

2 Department of Physics, University of Warwick, Coventry, CV4 7AL, UK

3 National Institute of Information and Communications Technology, Japan

### Status

Since the 1990s, the development of optical sources and detectors has accelerated in terahertz (THz) frequency region. Biomedical applications of terahertz spectroscopic and imaging systems using these technologies have benefited from increased radiance and sensitivity and have been applied to tissue assessment via the probing tissue water content, tissue birefringence and high order structure of protein. (Pickwell and Wallace, 2006; Taylor *et al.*, 2015; Markelz and Mittleman, 2022). Small changes in tissue hydration and protein structure often lead to measurable variations in THz refractive index and absorption coefficient, making THz spectroscopy a promising tool for diagnosing e.g. protein denaturation and early-stage cancers. High tissue water content and scattering from turbid tissue properties limit *in vivo* penetration depth. Therefore, a significant body of current work is focused on the cornea (Ozheredov *et al.*, 2018) and skin. Research efforts have explored and advanced implementation, including system development, theoretical analysis, *in vitro* and *in vivo* experimental studies, and a small but growing body of human subject measurements (Hernandez-Cardoso *et al.*, 2022; Qi *et al.*, 2024; Young *et al.*, 2024). Despite these advancements, the clinical implementation of THz technology remains challenging.

### Current and future challenges

Historically, much of THz biomedical research has been performed with general purpose spectroscopy systems that are configured to maximize spectral coverage and dynamic range. However, clinical and biomedical utilization of terahertz waves necessitate systems engineered end-to-end specifically for a target application which takes into consideration both the expected contrast mechanisms and the possible *in vivo* surface topography. For example, detecting abnormal endothelial water content as related to corneal disease benefits from system optics matched to the near-spherical profile of human cornea and a spectral band that can probe the resonant-like backscatter due to distinct layer structure. Conversely, detection, monitoring, and management of early-stage skin cancer, psoriasis, and dermatitis should leverage systems with large illumination bandwidths to increase to maximize contrast. Skin features a large range of layer thickness and hydration gradient profiles which can limit contrast. Increased bandwidth can increase contrast and enable temporal separation of return pulses from dielectric window front and back surfaces which aid in calibration.

On the data processing side, the development of numerical correction and measurement techniques to compensate for clutter due to curved uneven topography is ongoing but challenging as terahertz system optics render detection systems highly directive and dense, and user-friendly focal plane arrays do not exist. Additionally, THz waves are highly susceptible to diffraction as corneal radii of curvature are on the order of a 10's of wavelengths, skin surface roughness is in the sub-wavelength regime, and optical apertures are typically only 10's of wavelength in diameter.

Finally, it is often difficult to correlate THz imaging and spectroscopy data with other related data sets such as visual appearance, layer structure and thickness, pressure, or any other information that could improve detection/diagnosis if it was taken into consideration with THz measurements. Improvements to multimodal fiducial markers and integration of adjunct imaging/sensing modalities into the THz system aperture may improve the clinical utility of THz data.

## Advances in science and technology to meet the challenges

Novel imaging methods have been developed for scanning the near-spherical surface of human cornea (Sung *et al.*, 2018; Virk *et al.*, 2023). The system in Fig. 5 (a) takes advantage of an off-axis parabolic mirror's focusing properties. As demonstrated by the ray-tracing diagram scanning a collimated beam transverse to the mirror's clear aperture plane produces an angular scan at the focal point. If the cornea is positioned such that its center of curvature is coincident with the mirror focal point, then the cornea surface can be scanned at normal incidence while keeping the source, detector, and target stationary.

Diffraction artifacts were observed at the edges of the eyeball in all images of the characterization targets, either increasing or decreasing the signal in areas where the known target reflectivity did not vary. Ongoing research is focused on mitigating diffraction effects by optimizing the aperture size. Similar diffraction artifacts were observed in a THz analysis of rabbit eyes used for millimeter-wave exposure experiments (Mizuno *et al.*, 2021), and its effects were suppressed by extracting specular reflection components using a time-domain method.

For the accurate diagnosis of skin with surface roughness, a portable THz handheld scanner was developed, as shown in Fig. 5 (b) (Hernandez-Serrano *et al.*, 2024). A femtosecond pulse train is split into two arms. One arm generates ultrabroadband THz radiation (emitter) and second arm is used as an optical gate at the detector. Data can be analyzed directly in the time domain or analyzed in the frequency domain via the Fourier Transform (Fig. 5 (b)). The human body's non-uniform topography prevents system optical design to a specific shape a priori (unlike cornea) and thus it is typical to use a low loss, dielectric window (e.g. quartz) to flatten the field as seen in Fig. 5 (b). This configuration produces two distinct reflections; one from the air–quartz interface (labeled as "A") and the other from the quartz–skin interface labeled as "B"). The refractive index and absorption coefficient of the skin were calculated from the reflection signals. To ensure consistent force application on the skin during the measurement, two force-sensitive resistors were integrated into the tip of the probe. Furthermore, these optical properties were derived from data measured at 55 and 60 s, when the air gap between the quartz and skin of each subject was minimal. Additionally, the occlusion effect (blocking the skin) stabilizes after ~30 seconds of contact. This measurement technique enables the quantitative evaluation of human skin hydration gradients and stratum corneum thickness, demonstrating its potential for noninvasive dermatological assessment.

For measurement of skin and eyes, significant recent work has focused on integrating optical coherence tomography (OCT) with THz spectroscopic imaging. OCT utilizes a low-coherence light source to perform tomography by cohering a tissue backscattered beam with a copy of itself. The system design and data analysis methodologies are very similar to THz Time Doman Spectroscopy (TDS) and thus form a natural adjunct. An example of OCT integration with THz system experiments is shown in Fig. 5(c) where OCT imagery confirms increased corneal thickness 1 day after 162 GHz irradiance at 360 mW/cm$^2$; which is ~ 16x more than current safety standards dictate. On the other hand, THz reflectance from the cornea decreased just after the 162 GHz irradiance due to the acute drying. The integrated analysis enables discussion that the acute drying and heat cause a trigger for corneal epithelium damage.

Concluding remarks

THz sensing represents a promising approach for overcoming current measurement limitations, offering noninvasive, high-sensitivity detection of tissue hydration and structural changes. Advances in imaging systems and data correction methods are expected to enable clinical translation of THz technology, ophthalmology and dermatology, in the near future.

To ensure human protection in these cases, experimental clarification and data accumulation, particularly in the eyes (Kojima *et al.*, 2020) and skin (Sasaki *et al.*, 2017), are critical issues. This experimental data is necessary to revise the current safety standards that the THz band spans. RF guidelines of the International Commission on Non-Ionizing Radiation Protection (ICNIRP, 2020) sources from low frequency up to 300 GHz while the guidelines on laser radiation (ICNIRP, 2013) covers sources from ultraviolet down to 300 GHz. These standards do not agree at 300 GHz and it is clear that the success of future clinical applications will rely on clarifying application and frequency specific safety margins in the THz band.

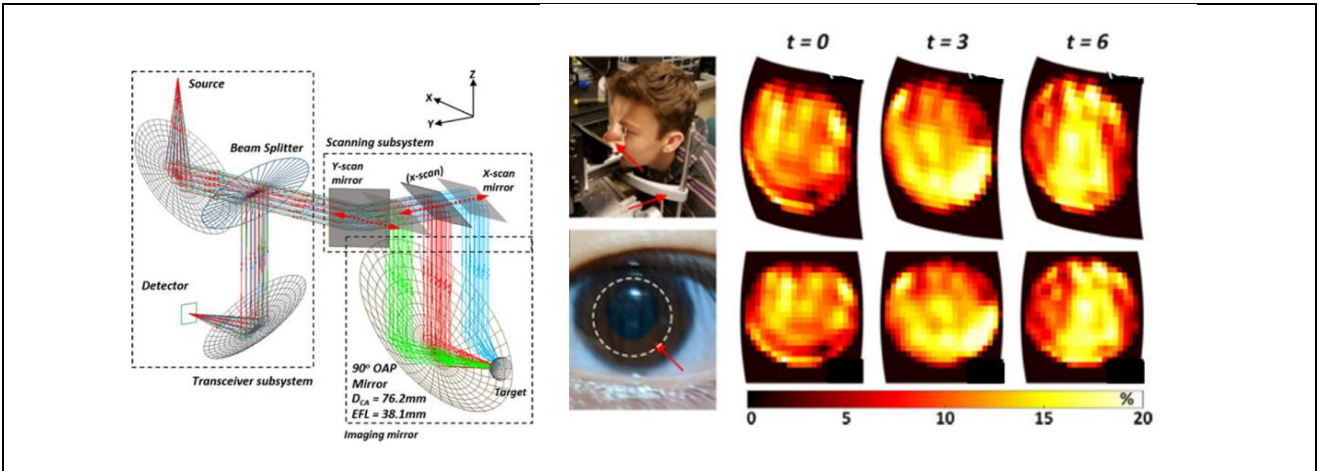

(a)

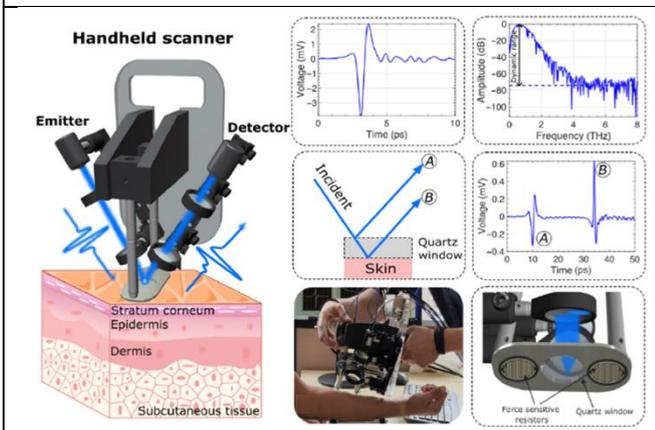

(b)

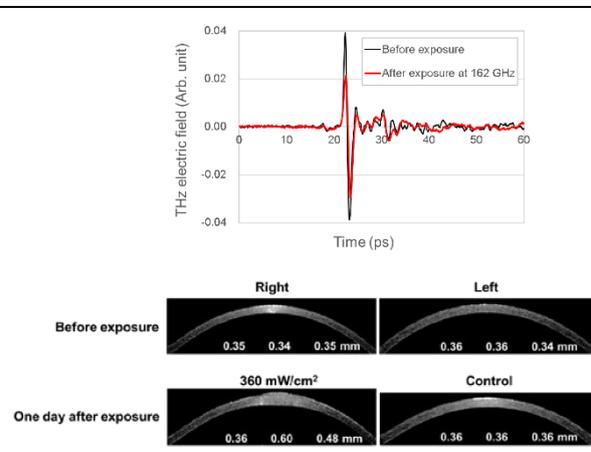

(c)

**Figure 5.** In vivo sensing systems and measurement examples for (a) human eyes and (b) skin (taken from Sung *et al.* 2018, Hernandez-Serrano *et al.* 2024), and (c) THz and OCT measurement examples for rabbit eyes (adapted with permission from Mizuno *et al.* 2021 © Optical Society of America).

# 6. Wireless Charging and Safety in Medical Devices

## 6.1. Wireless Charging of Implanted Medical Devices

Mauro Feliziani[1], Valerio De Santis[1], and Hubregt J. Visser[2]

1 Department of Industrial and Information Engineering and Economics, University of L'Aquila, Italy

2 Department of Electrical Engineering, Eindhoven University of Technology, The Netherlands

### Status

The global implantable medical devices (IMDs) market has been experiencing significant growth and is projected to expand further.

IMDs can be categorized by device type, with the most common ones being:

1. Orthopedic Implants: devices such as joint replacements, trauma fixation, and spinal implants;

2. Cardiovascular Implants: including pacemakers (PMs), implantable cardioverter-defibrillators (ICDs), cardiac resynchronization therapy (CRT) devices, and cardiac pumps/total artificial hearts;

3. Neurostimulators: devices such as Deep Brain Stimulators (DBS) that deliver electrical stimulation to specific parts of the nervous system;

4. Cochlear Implants: devices that provide a sense of sound to individuals with severe hearing loss;

5. Ophthalmic Implants: devices such as intraocular lenses replacing natural lenses in cataract surgery;

6. Capsules and Microrobots: although not implanted, these small ingestible devices with sensors and cameras travel through the digestive tract for diagnostic or therapeutic purposes.

Electric/electronic IMDs, generally known as active IMDs (AIMDs), are powered by an autonomous battery that must be replaced once exhausted. This needs surgical intervention for patients with potential health risks from complications (infections or failures).

Wireless Power Technology (WPT) has revolutionized AIMDs by enabling the transfer of electrical energy without the need for physical connectors (Kurs *et al.*, 2007; Jow and Ghovanloo, 2007; Si *et al.*, 2008; Ho *et al.*, 2013;

Agarwal *et al.*, 2017). Recent advancements in WPT for AIMDs has led to improving patient comfort, reducing infection risks, extending device longevity, and enabling AIMD functions that were previously inactive to save energy. These benefits motivated the transition from traditional power supplies to WPT systems.

The current electromagnetic WPT technologies for AIMDs can be classified based on various criteria:

- Number of transmitters (TX) and receivers (RX): the most common setup features a single external TX wirelessly powering an implanted RX. However, advanced configurations may incorporate multiple TXs and/or RXs to enhance the WPT performance.

- Operating Frequency: low-frequency (LF), intermediate-frequency (IF), and RF;

- Technology: near-, mid-, and far-field

- Coupling Mechanisms: capacitive, inductive, resonant inductive, radio-frequency/microwave, and optical coupling;

- Power Levels: ultra-low power, low power, and high power;

- Implantation Depth (range): deep, subcutaneous, and wearable;

- Powering System: battery-powered and battery-free implants.

For AIMD applications, the key indicators of WPT performance include power transfer efficiency (PTE), power level, range, size, and biocompatibility.

The typical performance of different WPT categories is summarized in Table I.

Table I: WPT Performance vs WPT Technology

| Feature | Near-Field WPT | Mid-Field WPT | Far-Field WPT |
|---------|----------------|---------------|---------------|
| Operating Principle | Inductive/Resonant Coupling | Evanescent/Quasi-Static Fields | RF/Microwave/Optical Waves |
| Range | Small (mm to cm) | Medium (cm to 10 cm) | High (10 cm to m) |
| Efficiency | High (~80-90%) | Moderate (~50-70%) | Low (~10-50%) |
| Power Level | High (mW to W) | Moderate (mW to W) | Low ($\mu$W to mW) |
| Application Suitability | Pacemakers, insulin pumps, neuro-stimulators | Deep-tissue implants, wireless biosensors | Smart implants, bioelectronic sensors |

This classification provides a clearer understanding of how different WPT technologies are applied in AIMDs with varying trade-offs in efficiency, range, power, and suitability for specific applications.

WPT applications have been investigated for common IMDs such as cochlear implants (Zeng *et al.*, 2008; Qian *et al.*, 2019), retinal implants (RamRakhyani and Lazzi, 2012; Weiland and Humayun, 2014; Mashhadi *et al.*, 2018), neuro-stimulators (Mirbozorgi *et al.*, 2017; Tanabe *et al.*, 2017; Barbruni *et al.*, 2020), pacemakers (Monti *et al.*, 2015; Campi *et al.*, 2016; Wang *et al.*, 2020a), cardiac defibrillators (Tang *et al.*, 2017; Campi *et al.*, 2021a), artificial heart, insulin pump (Ballardini *et al.*, 2023), drug pump (Rhee *et al.*, 2021), endoscopic capsules (Basar *et al.*, 2014; Campi *et al.*, 2023; Zhang *et al.*, 2024). Although IMDs with WPT technology are currently under development and experimentation worldwide, they are not already widely available in the market, even though some commercial applications exist, for example, Medtronics PERCEPT™ RC NEUROSTIMULATOR Rechargeable for DBS (https://www.medtronic.com/) and Cochlear Implants for Hearing Loss, MED-EL (https://www.medel.com/uk).

WPT technologies offer several advantages over the traditional solutions:

- Reduced Maintenance: AIMDs with rechargeable batteries can be wirelessly powered by an external source, eliminating the need for frequent battery or device replacements by surgical procedures.

- Improved Device Longevity: WPT allows for continuous power delivery and extends the operational life of the devices.

- Minimized Invasiveness: WPT eliminates the need for external cables and reduces the capacity, size, and weight of the internal battery, making it less expensive and easier to implant.

- Enhanced Patient Comfort: WPT eases the management of external power sources, improves patient comfort, and reduces complications.

- Safer Operation: WPT reduces the risk of mechanical failure and infection by eliminating external power connections.

- Advanced Functionality: WPT supports more advanced IMDs, enabling higher power and real-time data transmission with continuous power.

Owing to these advantages, WPT has the potential to improve the overall functionality, safety, and patient experience with IMDs, particularly in devices that require continuous operation inside the human body.

Challenges and Future Directions

Although the potential of WPT for IMDs is evident, several challenges remain in the development and widespread adoption of these technologies, including efficiency improvements, EMF safety, electromagnetic interference (EMI) mitigation, biocompatibility, standardization, and market acceptance. However, from a physical and biological perspective, the main problems mainly depend on the power requested from the AIMD and the depth of the implants (Poon *et al.*, 2010).

The power levels of AIMDs vary according to their functionality and design. Ultra-low-power devices, typically consuming less than 10 µW, include pacemakers and continuous monitoring sensors designed for long-term operation with minimal energy. Low-power devices, ranging from 10 μW to a few milliwatts, include implantable drug delivery systems, hearing implants, and cardiac devices that require low energy for specific tasks. High-power devices, which consume several milliwatts to watts, include neurostimulation devices, muscle stimulators, cardiac pumps, and microwave-based implants that require higher energy for therapeutic functions.

The electromagnetic challenges of wirelessly powering deep medical implants mainly involve the power transfer efficiency, EMF safety, and field penetration (Campi *et al.*, 2019). As the implant depth increases, energy transfer from the external source becomes less efficient owing to tissue attenuation, which increases with frequency. The need to ensure safe power transmission is critical, as higher power levels can cause adverse health effects such as tissue damage, neurostimulation, or overheating. Additionally, field penetration becomes more difficult at greater depths, and the surrounding tissue can absorb or scatter electromagnetic energy, reducing the efficiency of power transfer. Implant size has also become an issue, as deep implants require compact power receivers, which are harder to design for efficient deep power transfer. EMI with other devices and power transfer reliability can be problematic (Hikage *et al.*, 2016; Campi *et al.*, 2021b).

## Scientific and Technological Advances Addressing the Challenges

A range of scientific and technological advancements are helping to address the challenges of WPT in medical devices. However, the main challenge in applying WPT to implantable medical devices is the delivery of high power to deep implants, while ensuring compliance with EMF safety standards (Campi *et al.*, 2019; Lan *et al.*, 2025). To address this challenge, the following sections present three examples related to capsules, high-power implants, and future developments.

Electronic capsules rely on integrated batteries, but their limited capacity often restricts functions, such as drug release and electric stimulation. WPT offers a potential solution using a TX coil placed outside the human body and an RX coil inside the capsule. The power transfer efficiency depends on the capsule's position, with distances from the skin ranging from 2 cm to over 10 cm, a considerable distance given the attenuation of magnetic fields in biological tissues. The key design challenges include selecting the operational frequency, configuring the receiving coil, and designing the transmitting coil. The operational frequency must balance the field attenuation in biological tissues with the WPT performance. The receiving coil must be sufficiently small to fit the capsule (typically 2–2.5 cm) while ensuring adequate power transfer. A triaxial RX coil is more effective than a uniaxial RX coil because it can capture magnetic fields in/from any direction. However, because they are larger and heavier, research has focused on using an uniaxial RX coil in combination with a TX coil system capable of generating a spatially varying magnetic field in the human body. Various solutions have been proposed to address this issue (Zheng *et al.*, 2025; Campi *et al.*, 2023; Zhang *et al.*, 2024; Fadhel and Kamal, 2024).

One of the AIMDs characterized by high power consumption and deep implantation that could greatly benefit from WPT is the left ventricular assist device (LVAD), an advanced electrically powered mechanical pump that supports patients with severe heart failure. Currently, LVADs receive power externally via a percutaneous driveline (DL), which significantly increases the risk of severe infection at the exit site. LVADs require an average power of approximately 5 W, with peak demands reaching 20–25 W in the pulsatile mode. These devices are implanted directly attached to the heart, meaning that they are positioned several centimetres beneath the skin. As a result, wirelessly powering these implants using conventional architectures, where an external TX coil transfers energy to an internal RX coil mounted on the LVAD, presents significant challenges. To overcome these limitations, a simple yet highly efficient WPT architecture has been developed for implanted systems (Campi *et al.*, 2020). In this design, the RX coil is implanted subcutaneously, whereas the TX coil remains external, forming a well-established transcutaneous energy transfer (TET) coupler. The subcutaneous RX unit is connected to the LVAD through a fully internal DL. In addition, a rechargeable battery is implanted in the RX to provide supplemental energy. The internal battery supplies power during peak LVAD demands in pulsatile mode. Therefore, the WPT system is sized for average power consumption, whereas the battery compensates for short bursts of higher power. The WPT configuration with a battery allows patients to remain untethered for 1–2 h, significantly improving their mobility and quality of life. By eliminating the need for a constant connection, WPT-integrated LVADs offer greater freedom, reduced infection risk, and enhanced patient comfort (Campi *et al.*, 2021b; Liu *et al.*, 2018; Seshadri *et al.*, 2018; Cao and Tang, 2018).

Finally, only a few medical devices, such as artificial organs and cardiac pumps, require significant power; however, the demand for high-power AIMDs is expected to increase. Next-generation devices require continuous operation and substantial power consumption. Mainstream research typically focuses on powering each AIMD individually, with each AIMD equipped with its own WPT receiver. However, this approach is neither efficient nor safe when multiple AIMDs are implanted into the same patient. Recently, a new concept was proposed (Campi *et al.*, 2021a) that implants a wired network within the human body to link multiple AIMDs. This network can be powered externally by a high-power TET system that distributes energy to multiple AIMDs via wired connections. Additionally, the internal batteries of AIMDs can serve as shared backup power sources for all devices connected to the network. The proposed solution offers several benefits, including a reduced number of implanted WPT systems, higher efficiency, continuous high-power delivery, fewer surgical interventions for battery or device replacement, elimination of percutaneous cables, lower risk of infection, and the potential to deliver exceptionally high power to the human body, enabling novel and previously unimagined applications.

Concluding Remarks

WPT of medical devices is on the brink of revolutionizing healthcare by enabling the continuous operation of implanted devices without the need for invasive procedures or frequent battery replacement. This offers vast potential to improve the management of chronic medical conditions and enhance patients' quality of life.

Current technologies, such as inductive and resonant inductive coupling, have demonstrated promise in powering small- and medium-power medical devices; however, challenges related to power efficiency, EMF safety, and standardization still remain. Research and technological advances are addressing these challenges, particularly the development of more efficient energy transfer methods, improved biocompatibility, miniaturization, and creation of universal charging standards.

## 6.2. Safety for Implantable Medical Devices

Ji Chen

Department of Electrical and Computer Engineering, University of Houston, US

### Status

Implantable medical devices are increasingly vital in supporting various organ functions, thereby enhancing overall societal well-being (Bazaka and Jacob, 2012; Khanna, 2016; Wilson *et al.*, 2021; Magisetty and Park, 2022). Based on their operational characteristics (see Table 2), these devices can be classified into passive implantable medical devices (PIMD), which do not require a battery for functionality, and active implantable medical devices (AIMD), which rely on a power source or wireless charging to operate (European Parliament and Council, 2017). When patients with these devices undergo magnetic resonance imaging (MRI) scans, significant safety concerns arise—such as RF induced heating and gradient field-induced peripheral nerve stimulation (PNS) (Schueler *et al.*, 1999; Nyenhuis *et al.*, 2005; Woods, 2007; Dal Molin and Hecker, 2013; Davids *et al.*, 2019; Jabehdar Maralani *et al.*, 2020; Wang *et al.*, 2022c; Yang *et al.*, 2025). Typically, AIMD systems include devices such as pacemakers, various neuromodulation systems (e.g., spinal cord stimulators, deep brain stimulators), cochlear implants, and brain-machine interface systems. In contrast, PIMD systems often comprise stents, orthopaedic plates, pedicle screw systems, hip replacements, knee replacements, and similar devices.

Table 2. Key Features of Passive and Active Implantable Devices.

| Feature | Passive Implantable Medical Devices (PIMDs) | Active Implantable Medical Devices (AIMDs) |
| --- | --- | --- |
| Power Source | No external power required | Requires an internal/external power source |
| Functionality | Primarily structural or mechanical | Electronic or electromechanical |

| Interaction with Body | Limited direct interaction, relies on body mechanics | Actively regulates or influences body functions |
|---|---|---|

To ensure patient safety, regulatory bodies have developed guidance documents that integrates international standards to address these concerns. For all AIMDs, the following MRI evaluations should be performed (U. S. FDA, 2023; Lidgate *et al.*, 2024; ASTM F2052, 2022; ASTM F2119, 2024; ASTM F2182, 2020; ASTM F2213, 2017; ISO/TS 10974, 2018):

1. RF-induced heating in the patient
2. Gradient-induced device heating and RF-induced device heating in the patient
3. Gradient-induced vibration in the patient and device
4. B0-induced force in the patient
5. B0-induced torque in the patient
6. Gradient-induced extrinsic electric potential in the patient
7. B0-induced malfunction in the device
8. RF-induced malfunction in the device and RF rectification in the patient
9. Gradient-induced malfunction in the device
10. Combined fields test for both patient and device
11. Image artifacts

For all PIMDs, the required evaluations are items 1, 4, 5, and 11 from the list above.

Based on the results of these tests, appropriate MRI conditions can be established to safely scan patients with implantable devices.

Challenges and Future Directions

The safety of implantable devices under MRI exposure can be characterized as a standard electromagnetic compatibility (EMC) problem, with challenges arising from the EMC source, coupling pathways to the implant, and the device's immunity to external electromagnetic emissions. Likewise, solutions also lie within these domains. Despite advancements in current testing methods to address these issues, significant challenges persist.

1. Large uncertainties exist in the RF safety assessment.

The state-of-the-art method for AIMD RF related evaluations involve the device model development using the transfer function (TF) method as described in the ISO/TS 10974 (2018)g. This procedure requires one to use both numerical modelling and experimental measurement to estimate the RF related heating near the device tip. Due to the variations in human body type, different implantation pathways inside human body, tissue properties near implantation pathways, measurement equipment accuracy, and numerical modelling precision, there is a large uncertainty budget typically around 25% needs to be accessed on top of the results (Yao *et al.*, 2018; Zeng *et al.*, 2018a; Wang *et al.*, 2020b; Winter *et al.*, 2021). Consequently, the developed conditions for safe scan of patients are often very conservative and can prohibit patients from scanning under some aggressive and high-resolution MR sequence.

2. Ultra-high field systems 5T, 7T and beyond

With the development of whole-body 5T and head-imaging 7T systems, the current methodology on device evaluation under 1.5T and 3T needs to be revised and updated (Barisano *et al.*, 2019; Fagan *et al.*, 2020). It was observed that the conventional method of using TF models to characterize device behaviour under 3T is no longer applicable for high fields system over 7T and 10.4T (Abedi *et al.*, 2022; Zuo, 2025). For ultra-high field system, the safety evaluation may require more rigorous method, such as full-wave electromagnetic characterization tools.

3. Fast switching dB/dt lead to device neuron stimulation

For MR system related safety concerns, the RF-induced heating is typically the bottle neck that restricts the condition for safe MR scans. However, as the fast gradient switch of dB/dt increases, the unintended PNS stimulation will have pronounced effect due to the implants. Conventionally, only the AIMD needs to consider the gradient induced intended stimulation, however, recently a study found out that large PIMD can also lead unintended nerve stimulation near the ends of these devices (Yang *et al.*, 2025).

4. Implant interactions with other medical equipment, such as RF ablation, electrical surgical knife, and transcutaneous electrical nerve stimulation systems (TENS).

With the operation of nearby RF ablation system, electrical surgical knife, and TENS systems, due to the antenna effect of the implantable devices (both AIMD and PIMD), the implantable devices can attract surrounding electromagnetic energy and redistribute these energies based on the factors such as shape and insulation thickness around them. Depending on the operation frequency of the surrounding medical equipment, such energy redistribution can lead to significant heating concerns or neural stimulation (Pfeiffer *et al.*, 1995; Holmgren *et al.*, 2008; Paniccia *et al.*, 2014).

Scientific and Technological Advances Addressing the Challenges

1. Large uncertainties in the MRI RF safety assessment can potentially be addressed through patient specific approach in the safety evaluation. With the artificial intelligence (AI) assisted segmentation method, it is feasible to quickly develop a patient model using low energy MR scan or X-ray method. Based on the developed human body model and pathway identified inside the human body model, one can quickly access the patient specific MR safety, especially the RF related safety concerns. With more and more human models developed using this approach, a large AI based imaging library can quickly assess the MR safety for various implantable devices (Aissani *et al.*, 2019; Zheng *et al.*, 2021; Vu *et al.*, 2021; Chang *et al.*, 2022).

2. Ultra-high field systems, such as 5T, 7T, and beyond, pose significant challenges but also offer innovative solutions. These systems typically require multichannel methods to achieve $B_1$ field homogeneity for imaging purposes. However, this multichannel approach can also be leveraged to enhance safety in the presence of implants. For instance, a multichannel high-field system can be optimized to maintain field homogeneity in the region of imaging interest while simultaneously minimizing emission power in areas where an implanted device is located, reducing potential safety risks (van den Bergen *et al.*, 2009; Tang *et al.*, 2011; Zeng *et al.*, 2018b; Aberra *et al.*, 2020b).

3. Fast-switching dB/dt can lead to neuronal stimulation, which can be better addressed through more accurate neuron modelling. Traditionally, safety limits have been established based on induced charge limits, derived from experimental studies. While this approach is conservative to ensure patient safety, it may not fully capture real physiological responses. With advances in electromagnetic modelling and the use of the NEURON code, it is now possible to conduct patient-specific electromagnetic simulations. These simulations can provide a realistic electric field distribution along nerve fibres, which can then be integrated into the NEURON modelling tool. This approach enables the identification of actual neuronal activation rather than relying solely on the accumulated charge limit, leading to more precise and individualized safety assessments (Carnevale and Hines, 2006; Han, 2018; Tan *et al.*, 2020; Zilberti *et al.*, 2024).

4. Interactions between implants and other medical equipment, such as RF ablation, electrosurgical knives, and TENS, pose significant safety concerns. However, with recent advancements in multiphysics and multi-scale modelling, neuron physiology, and GPU/cloud-based computational power, it is now feasible to conduct in-silico studies to better understand and mitigate these interactions. These advanced simulations enable more accurate assessments, thereby improving both device safety and treatment efficacy (Han, 2018; Akbari, 2021; Kenjereš, 2014).

Concluding Remarks

While significant advancements have been made in assessing MR safety for patients with medical implants, the growing landscape of other medical equipment—such as TENS devices and RF ablation systems—introduces new challenges for implant safety. Additionally, many everyday applications involve electromagnetic sources that may emit unintended signals. While these emissions are generally considered safe for individuals without implants, they can pose potential risks for patients with implanted medical devices. Examples include induction cooktops, electric vehicles, and various electronic systems. To ensure patient safety, it is imperative to thoroughly investigate the interactions between implants and various electromagnetic emission sources. A comprehensive understanding of these interactions will help mitigate risks and improve safety standards for individuals with medical implants.

## 6.3. Pacemaker Malfunction: Interference of Electromagnetic Fields

Takashi Hikage[1] and Seungyoung Ahn[2]

1: Faculty of Information Science and Technology, Hokkaido University, Japan

2: Cho Chun Shik Graduate School of Mobility, Korea Advanced Institute of Science and Technology, Daejeon, South Korea

### Status

Pacemakers are critical medical devices that regulate cardiac rhythms and provide life-saving support to patients with arrhythmias or other heart conditions. However, the functionality of pacemakers can be compromised by exposure to EMFs, which are becoming increasingly ubiquitous due to the proliferation of electronic devices and communication systems. Numerous studies have identified specific frequency ranges and intensities of EMFs that can interfere with pacemaker operations and cause malfunctions, such as inappropriate pacing, inhibition, and mode switching (Barbaro *et al.*, 1995; Irnich *et al.*, 1996; Toyoshima, 1996; Hayes *et al.*, 1997; Wang *et al.*, 2000; Grant and Schlegel, 2000; Mattei *et al.*, 2014). Regulatory bodies, including the International Commission on Non-ionizing Radiation Protection and Institute of Electrical and Electronics Engineers, have established guidelines (IEEE-C95.1, 2019; ICNIRP, 2020) for EMF exposure to minimize risks; however, real-world incidents highlight persistent vulnerabilities near devices, such as Electronic Article Surveillance and Wireless Power Transfer systems that generate strong magnetic fields in relatively low-frequency bands, where electromagnetic interference effects can occur (ISO/IEC TR 20017, 2011; Hikage *et al.*, 2016; Driessen *et al.*, 2019; Campi *et al.*, 2021b; Maradei *et al.*, 2024). Furthermore, particularly in environments with high-intensity or unexpected EMF exposure, such as industrial zones, during medical imaging procedures, such as magnetic resonance imaging (MRI), or near devices that generate strong magnetic fields, the potential impact of metal heating should not be overlooked.

### Challenges and Future Directions

Figure 6 illustrates the key challenges and future directions for addressing electromagnetic interference (EMI) risks for pacemakers. Despite considerable advancements in pacemaker technologies, several challenges remain.

1. **Increased EMF Sources**: The growing deployment of 5G networks, wireless power transfer systems, and wearable devices has increased the likelihood of exposure to diverse EMFs. Because individuals with pacemakers are likely to come into the close proximity to these devices, careful investigation and research are essential.

2. **Comprehensive Testing Standards**: Current standards for pacemaker testing under EMF exposure may not fully replicate real-world scenarios, leading to gaps in safety assessments (ISO/IEC 14117d, 2012; ISO 14708-2, 2019; ANSI/AAMI PC69, 2000; ANSI/AAMI PC69, 2007; ISO 14708-6, 2019). The development of testing methods that address various frequencies, signal waveforms, and applications, along with achieving global standardization, remains a critical challenge.

3. **Complex Interactions**: Pacemaker susceptibility to EMFs varies according to device design, patient anatomy, and environmental factors, complicating the development of universal mitigation strategies.

4. **Enhancing Risk Communication:** Further promotion of risk communication is essential. These efforts should target not only pacemaker users but also electronic device users, thereby fostering a broader understanding of pacemaker EMI-related risks.

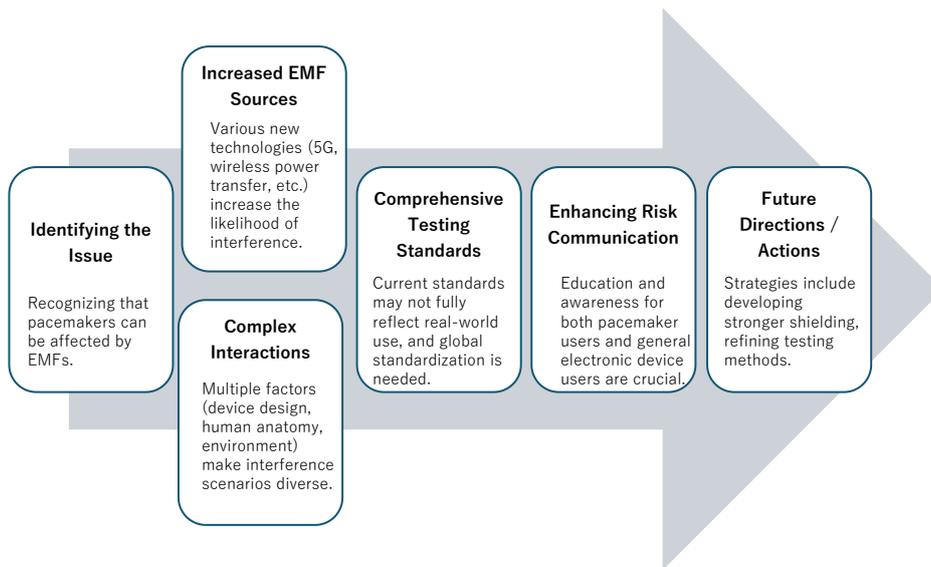

Fig. 6. Impact of EMF on pacemakers.

Future research should focus on developing more robust EMF shielding technologies for pacemakers, thereby enhancing device programming to detect and mitigate EMF interference and establishing more rigorous and realistic testing protocols. Interdisciplinary collaboration among medical device manufacturers, regulatory bodies, and researchers is essential to effectively address these challenges.

Scientific and Technological Advances Addressing the Challenges

Recent scientific and technological advancements offer promising solutions to mitigate EMF-related risks for pacemaker users.

1. **Enhanced Shielding and Filtering**: Advances in materials and technologies have improved the resistance of pacemakers to high-frequency EMFs, particularly those emitted by modern communication devices, such as mobile phones and cellular devices (Anjitha and Sunitha, 2024).

2. **MRI-Compatible Devices**: Innovations in MRI-safe pacemakers and leads have reduced the risks associated with one of the most intense sources of EMF exposure in clinical settings (Mitka, 2011; Indik *et al.*, 2017).

3. **Simulation-Based Testing**: Sophisticated computational models of the interactions between human anatomy and EMFs can accurately assess pacemaker behaviour under various exposure conditions. These models contributed to the development of safer device designs (Hikage *et al.*, 2020; Waki *et al.*, 2023). Nevertheless, the development of methods for verifying the accuracy and validity of the simulation results remains an important challenge.

Concluding Remarks

Although significant strides have been made in addressing the effects of EMFs on pacemaker functionality, challenges persist due to the rapidly evolving EMF environment and the diverse needs of pacemaker users. Future research should prioritize multidisciplinary approaches that integrate technological innovation, patient education, regulatory updates, and the promotion of effective risk communication to ensure reliability and safety of pacemakers in increasingly complex electromagnetic environments. Addressing these challenges will contribute to improving the quality of life of pacemaker users and maintaining their safety in a technology-driven world.